%% file: ms5.tex
\documentclass[12pt,preprint]{aastex}
\usepackage{amsmath}

\newcommand{\bdv}[1]{\mbox{\boldmath$#1$}}

\def\au{{\rm au}}

\def\masyr{{\rm mas}\,{\rm yr}^{-1}}
\def\kpc{{\rm kpc}}
\def\mas{{\rm mas}}

\def\muas{\mu{\rm as}}

\def\max{{\rm max}}

\def\rel{{\rm rel}}

\def\eff{{\rm eff}}

\def\e{{\rm E}}
\def\bpi{{\bdv\pi}}
\def\bmu{{\bdv\mu}}

\def\bgamma{{\bdv\gamma}}

\begin{document}
\title{Mass Production of 2021 KMTNet Microlensing Planets I}

\input author.tex

\begin{abstract}

We inaugurate a program of ``mass production'' of microlensing planets
discovered in 2021 KMTNet data, with the aim of laying the basis for
future statistical studies.  While we ultimately plan to quickly 
publish all 2021 planets meeting some minimal criteria,
the current sample of four was chosen
simply on the basis of having low initial estimates of the planet-host
mass ratio, $q$. It is therefore notable that 2 members of this sample suffer from
a degeneracy in the normalized source radius $\rho$ that arises from
different morphologies of closely spaced caustics.  All four 
planets [KMT-2021-BLG-(1391,1253,1372,0748)] have well-characterized 
mass ratios, $q$, and therefore are suitable for mass-ratio frequency
studies. Both of the $\rho$ degeneracies can be resolved by future 
adaptive optics (AO) observations on 30m class telescopes.
We provide general guidance for such AO observations for all events
in anticipation of the prospect that they will revolutionize the field
of microlensing planets.

\end{abstract}


\section{{Introduction}
\label{sec:intro}}

The rate of microlensing
planet detection has increased rapidly in recent years.  For 2021,
we estimate that of order 40 planets may be discovered that have
the Korea Microlensing Telescope Network (KMTNet, \citealt{kmtnet}) 
contributing a major part, or all, of the data underlying these
detections.  A substantial minority of these detections can lead to
individual-event publications, either because of the complexity of
the analysis, or the scientific importance of the planet (or planetary system).
In other cases, individual-event papers can be an important point of
entry of students and postdocs into the field.

However, for the great majority of planetary events currently being 
discovered, the main scientific interest is that they contribute
to the statistical sample of microlensing planets, which can then be exploited
to learn about the population as a whole.  At first sight, it may appear that 
it is not strictly necessary to publish the analysis of all of these events.
As an alternative, one might simply list the planets in a publication
devoted to the first statistical analysis that referenced them.
However, at least at the present stage, it is actually necessary to analyze 
these planetary events at a similar level to that required for publication,
and also to document these analyses in publicly accessible form.
First, without such analysis and documentation, it would be difficult
for other researchers to conduct their own statistical investigations,
possibly using alternative selection criteria.  Second, it is very likely
that a decade from now, with the inauguration of adaptive optics (AO)
observations on 30m class (``extremely large'', ELT) 
telescopes, essentially all hosts of 
microlensing planets that have been discovered (or will be discovered
in the next few years) will be directly imaged.  When this imaging
is combined with the original analyses, the masses, distances, and
planet-host projected separations will be determined with good precision.
In combination with a sound understanding of the selection function
from statistical studies, this additional information will revolutionize
the field.  However, without direct access to the original analyses
and data, the difficulty of incorporating late-time imaging would
be greatly increased.

The most straightforward and secure way to provide these analyses and
data is to publish all of these planetary events.  The experience of
\citet{kb190253} suggests that of order five planetary events can suitably
be grouped in one paper.  On the one hand, this allows various ``routine''
information, such as field position, cadence, source color, etc., to be
grouped into tables, rather than expressed as repeated narrative.  Moreover,
various procedures and formulae only need to be presented once per paper.
On the other hand, grouping planetary events into papers does not change
the amount of work required for the analysis, nor does it change the
amount of space required for exposition of the particular details for
each event.  Therefore, we begin, the ``mass production'' of 2021
planetary events with this four-event paper.

At present, KMTNet planets are discovered in three channels.  The main
traditional channel has been to identify potentially planetary events by eye 
using publicly available data from the KMTNet website, determine whether there
are corroborating (or contradicting) data from other surveys 
(i.e., the Optical Gravitational Lensing Experiment (OGLE) and
the Microlensing Observations in Astrophysics (MOA) surveys), and then
proceed with a more detailed analysis.  This leads to re-reduction
of the data when events prove to be of sufficient interest.  

The second channel is to densely monitor known KMTNet events (typically with
narrow-field telescopes, usually ranging from 25cm to 1m), and search
for planets in the combined survey(s) plus followup data.  The search
methods are similar to the first channel, but the data sets are by
nature idiosyncratic.  While the survey+followup approach was the most common
channel for planet discovery in early years (e.g., the second microlensing
planet, OGLE-2005-BLG-071Lb, \citealt{ob05071}), it was generally not
applied by KMTNet prior to 2020.  The major exception is that microlensing
events that were targeted for {\it Spitzer} observations 
\citep{yee15} were usually subjected to follow-up campaigns.  For example,
the mass-ratio $q\sim 10^{-5}$ planet in the {\it Spitzer} target 
OGLE-2019-BLG-0960 shows up most dramatically in follow-up data from
an amateur class telescope \citep{ob190960}.  However, in 2020, KMTNet began to
actively collaborate with the Microlensing Follow Up Network ($\mu$FUN)
and Tsinghua Microlensing Group to densely monitor high-magnification
events, which immediately led to the discovery of another $q\sim 10^{-5}$
planet, KMT-2020-BLG-0414Lb \citep{kb200414}.

The third channel is comprised of planets discovered using the 
KMT AnomalyFinder \citep{ob191053}, which has so far been applied only
to the $13\,{\rm deg}^2$ of KMT fields with cadence $\Gamma\geq 2\,{\rm hr}^{-1}$
and only to 2018-2019 data.  In fact, only eight of the newly discovered
planets from this search have been reported, including seven with
mass ratios $q\leq 2\times 10^{-4}$ \citep{ob191053,kb190253}, and one
wide-orbit planet \citep{ob180383}.  However, there are expected to be
many additional new planets when the AnomalyFinder is applied to additional
seasons, to lower-cadence fields, and when the $q>2\times 10^{-4}$ planets
are thoroughly investigated.

It is important to systematically analyze the planets from all three channels.
The third channel (AnomalyFinder) can be directly compared to a
planet-detection efficiency analysis (Y.K.~Jung et al., in prep) to yield
the planet-host mass-ratio function, and potentially the distribution
of projected separations.  The second channel (survey+followup) can
also be subjected to statistical analysis despite its seemingly chaotic
selection process \citep{gould10}.  While the first channel (by-eye selection)
cannot itself be subjected to rigorous statistical analysis, it serves
as an important external check on the AnomalyFinder selection process.
For example, \citet{kb190253} used by-eye detections to identify
three planets that were missed by the AnomalyFinder selection.
All three ``failures'' were explained by known effects, i.e., one
was below the $\chi^2$ threshold, one was in a binary-star system,
and one was ``buried'' in a high-magnification event with strong finite-source
effects at peak.  The latter two ``failures'' imply that the AnomalyFinder
program requires additional steps to find all planets.  It is important to
continue this vetting of the AnomalyFinder.  

Of course, there is substantial overlap between these three channels.
For example, \citet{kb190253} showed that 23 of the 30 planets that they
reported were already known, most being already published but with some in
preparation.  And some planets that are discovered in real time
by survey+followup will have sufficient survey data that they will
later be rediscovered by the AnomalyFinder search.  OGLE-2019-BLG-0960Lb
\citep{ob190960}, which was mentioned above, is a good example.  For
planets that are discovered through two channels, it does not matter
which channel is reported first, except that for survey+followup planets
that are recovered by AnomalyFinder, the generally larger error bars
(and possible increase in discrete degeneracies)
for the latter must be reported.

Here we begin the systematic publication of all 2021 planets
that were discovered by eye, as a component of this ``mass production'' 
approach.  As of this writing, there are (based on preliminary analysis) 
36 planet-candidate signatures\footnote{Some of these candidates will
not survive the detailed vetting and analysis leading to publication.
For example, in the course of deriving the sample of 4 planets for the
present paper, we had to investigate a total of 7 candidates, thereby
eliminating 3 of these.
KMT-2021-BLG-0637 was eliminated because re-reduction showed that the
apparent anomaly had been due to data artifacts. 
KMT-2021-BLG-0750 was eliminated because, after re-reduction, it was
preferred over a point lens by only $\Delta\chi^2=10$.
KMT-2021-BLG-0278 was eliminated because a binary-source solution was
preferred at $\Delta\chi^2=31$.  In contrast to the other two, this
elimination occurred when the paper was close to completion: originally
the paper would have reported 5 planets.}
that were discovered by eye from survey data
in a total 33 events from 2021, plus an additional 8 planet signatures
from 7 events that were discovered from events with followup observations.
We restrict consideration to the former.  Some of these, for example,
the three two-planet events, will be subjected to individual analysis.
Some others will be grouped together according to scientific themes.
We organize the remainder into groups by convenience.  The first group,
analyzed here, consists of the four events with planets of the 
lowest $q$ among those identified by YHR, according to the preliminary analysis 
(KMT-2021-BLG-1391, 
KMT-2021-BLG-1253, 
KMT-2021-BLG-1372, 
KMT-2021-BLG-0748). 

\section{{Observations}
\label{sec:obs}}

All four planets described in this paper were identified in by-eye
searches of KMT events that were announced by the KMT AlertFinder
\citep{alertfinder} as the 2021 season progressed.
KMTNet observes
from three identical 1.6m telescopes, each equipped with
a $(2^\circ\times 2^\circ)$ camera at CTIO in Chile (KMTC),
SAAO in South Africa (KMTS), and SSO in Australia (KMTA).
KMTNet observes primarily in the $I$ band, with 60 second exposures.
After every tenth $I$-band observation, there is a 90 second $V$-band
exposure (except for a small subset of observations taken as the
western fields are rising or the eastern fields are setting).
The data were reduced using pySIS \citep{albrow09}, which is a form of
difference image analysis (DIA,\citealt{tomaney96,alard98}).
Although all the planets were identified during the season using online
photometry, all of the light curves were re-reduced using the tender-loving care
(TLC) version of pySIS.  In particular, the algorithms and procedures of
this TLC pySIS are the same as has been applied in the AnomalyFinder papers
that were discussed in Section~\ref{sec:intro}.
For each event, we manually examined the images during the anomaly
to rule out image artifacts as a potential explanation for the
light-curve deviations. None of these four events were alerted by any other survey. 
To the best of our knowledge, there were no follow-up observations.

We follow the example of \citet{kb190253} by presenting a summary of
observational information for the four events in tabular form.
Table~\ref{tab:names} gives the event names, 
observational cadences $\Gamma$, as well as the discovery
dates and the sky locations.

\section{{Light Curve Analysis}
\label{sec:anal}}

We follow \citet{kb190253} by first presenting some procedures
and methods of analysis that are common to all events as a ``preamble''.
We refer the reader to that paper for some details that we
do not recapitulate here.

{\subsection{{Preamble}
\label{sec:anal-preamble}}

Most planetary microlensing events present themselves as basically 1L1S, i.e., 
\citet{pac86} light curves with short-term anomalies.  Here,
$n$L$m$S means $n$ lenses and $m$ sources.  These are characterized by three
parameters (in addition to two flux parameters for each observatory):
$(t_0,u_0,t_\e)$, i.e., the time of lens-source closest approach,
the impact parameter (in units of the Einstein radius, $\theta_\e$),
and the Einstein timescale,
\begin{equation}
t_\e = {\theta_\e\over\mu_\rel}; \qquad
\theta_\e\equiv\sqrt{\kappa M\pi_\rel};\qquad 
\qquad \kappa\equiv {4 G\over c^2\au}\simeq 8.14\,{\mas\over M_\odot}.
\label{eqn:thetaedef}
\end{equation}
Here, $M$ is the lens mass, $(\pi_\rel,\bmu_\rel)$ are the
lens-source relative (parallax, proper motion), and $\mu_\rel=|\bmu_\rel|$.
Four additional parameters are generally required to describe the
planetary perturbation: $(s,q,\alpha,\rho)$, i.e., the planet-host
separation (in units of $\theta_\e$), the planet-host mass ratio,
the angle between the source trajectory and planet-host axis, and
the angular source size $\theta_*$ normalized to $\theta_\e$.

Often the anomaly can be localized as taking place at $t_{\rm anom}$,
with duration $\Delta t_{\rm anom}$.  The anomaly is then offset from
the peak by an amount (scaled to $\theta_\e$), 
$\tau_{\rm anom} = \Delta t_{\rm anom}/t_\e = 
(t_{\rm anom} - t_0)/t_\e$.  At this point, the source
is separated from the host by $\Delta u_{\rm anom} = \sqrt{\tau_{\rm anom}^2+u_0^2}$.

If the anomaly is due to the source crossing a planetary caustic,
then $s=s_\pm^\dagger$ and 
\begin{equation}
s^\dagger_\pm \equiv {\sqrt{u_{\rm anom}^2 + 4} \pm u_{\rm anom}\over 2};
\qquad
\tan\alpha = +{u_0\over\tau_{\rm anom}},
\label{eqn:tau_anom}
\end{equation}
where the ``$\pm$'' refers to major-image and minor-image crossings,
respectively \citep{gouldloeb}. Note that $s^\dagger_- = 1/s^\dagger_+$ and 
$u_{\rm anom} = s^\dagger_+ - s^\dagger_-$.  Also note the sign difference
for $\tan\alpha$ relative to \citet{kb190253} due to different sign
conventions of the underlying fitting programs.
\citet{kb190253} argued that if the source does not
cross the planetary caustic but comes close enough to generate an
anomaly, then there can be an ``inner/outer degeneracy'' \citep{gaudi97},
for which $s^\dagger$ is the arithmetic mean of the two solutions.
They showed that this was an excellent approximation for the four cases
that they presented.
Note that for non-caustic-crossing
anomalies, major-image perturbations usually appear as a ``bump'',
while minor-image perturbations appear as a ``dip''.  
In the course of applying this framework to a larger sample of planets,
Gould et al. (2022, in prep), found that $s^\dagger$ was a better approximation
to the geometric than the arithmetic mean of the two solutions,
$s^\dagger \rightarrow \sqrt{s_{\rm inner} s_{\rm outer}}$,
which immediately led to a unification of the inner/outer degeneracy
for planetary caustics \citep{gaudi97} with the close/wide degeneracy
for central and resonant caustics \citep{griest98}, as conjectured by 
\citet{ob190960}.  This conjecture was motivated by the gradual accumulation
of events for which the ``outer'' degenerate solution is 
characterized by the source passing outside the planetary wing of a resonant
caustic rather than an isolated planetary caustic \citep{ob180677}.
We refer the reader to Gould et al. (2022, in prep) for the genesis of
this discovery.  Here we focus on providing homogeneous notation for the
unified $s^\dagger$ formalism.

First, we write the theoretical (``heuristic'') prediction for $s_\pm^\dagger$
as given by Equation~(\ref{eqn:tau_anom}), and including the ``$\pm$''
subscript according to whether the anomaly appears to be a major-image
or minor-image perturbation.  Then we define $s^\dagger$ (without subscript)
to be the result of 
combining two empirically derived related solutions, $s_+ > s_-$, i.e., 
\begin{equation}
s^\dagger=\sqrt{s_+ s_-}.  
\label{eqn:s_geomean}
\end{equation}
The prediction ($s^\dagger_\pm$) 
can then be directly compared to the empirical result ($s^\dagger$).
The relation between $s_\pm$ and $s^\dagger$  can also be expressed as
\begin{equation}
s_\pm = s^\dagger\exp(\pm\Delta\ln s)
\label{eqn:s_pm}
\end{equation}
where $\Delta\ln s\equiv (1/2)\ln{s_+/s_-}$.  This replaces and generalizes the
formalism of \citet{kb190253}, for whom, effectively
$s_\pm = s^\dagger\pm \Delta s$.  Using this formalism, we can unify the
three regimes of major-image planetary caustic, central/resonant caustic, and 
minor-image planetary caustic, by identifying
\begin{align}
s_{\rm inner}&\rightarrow s_+,  \quad
s_{\rm outer}\rightarrow s_-, \quad ({\rm major\ image},\ s^\dagger>1)\\
s_{\rm wide}&\rightarrow s_+, \quad
s_{\rm close}\rightarrow s_-,  \quad ({\rm central/resonant}, s^\dagger\simeq 1) \\
s_{\rm outer}&\rightarrow s_+, \quad
s_{\rm inner}\rightarrow s_-, \quad ({\rm minor\ image},s^\dagger <1).
\end{align}
Note that, as discussed by \citet{ob190960}, the $s_\pm$ degeneracy
is actually a continuum spanning from 
$s_+^\dagger>1$ (major image caustic) through
$s_\pm^\dagger\simeq1$ (central/resonant caustic) to
$s_-^\dagger<1$ (minor image caustic), and it may be difficult
to distinguish between ``close/wide'' and ``inner/outer'' designations
in any particular case because these designations are simply nodal points
on the continuum.
We will consistently employ this framework and notation in the current work.
See \citet{zhang22} for another approach to unification.

\citet{kb190253} also showed that for minor-image perturbations,
the mass ratio could be estimated
\begin{equation}
q = \biggl({\Delta t_{\rm dip}\over 4\, t_\e}\biggr)^2
{s\sin^2\alpha\over u_{\rm anom}} 
= \biggl({\Delta t_{\rm dip}\over 4\, t_\e}\biggr)^2
{s\over |u_0|}|\sin^3\alpha| .
\label{eqn:qeval}
\end{equation}
We note here that this expression can be rewritten in terms of
``direct observables'' as,
\begin{equation}
t_q \equiv q t_\e = {1\over 16}{(\Delta t_{\rm dip})^2\over t_\eff}
\biggl(1 + {(\delta t_{\rm anom})^2\over t_\eff^2}\biggr)^{-3/2}s_\pm^\dagger,
\label{eqn:qeval2}
\end{equation}
where we have substituted $s\rightarrow s_\pm^\dagger$ because it is this
quantity that can be estimated by eye.
That is, $\Delta t_{\rm dip}$ and $\delta t_{\rm anom}$ can be read directly
off the light curve, while $t_\eff\simeq {\rm FWHM}/\sqrt{12}$ for even
moderately high magnification events, $A_\max \ga 5$.  Typically for such events,
$u_{\rm anom}\ll 1$, implying $s^\dagger\rightarrow 1$.  Hence, the right hand
side is an invariant, i.e., independent of $t_\e$, which can be difficult
to measure accurately in some cases.  Thus, $t_q=q t_\e$ is also an
invariant \citep{mb11293}.

In the six cases that \citet{kb190253} presented, 
Equation~(\ref{eqn:qeval}) proved accurate
to within a factor 2, with the main problem being the difficulty
of estimating $\Delta t_{\rm dip}$, which enters quadratically,
from the light curve.  Based on the work of \citet{chung11}, they
expected this approximation to deteriorate for $q\ga 2\times 10^{-4}$.
However, for the events examined here (as in \citealt{kb190253}), $q$ is
close to or below this approximate boundary.

However, even when the anomaly appears to be isolated, it is not necessarily
due to a small planetary caustic.  It can, for example, be due to a cusp
crossing or cusp approach of a much larger caustic.  In such cases,
the $(s,\alpha)$ predictions of Equation~(\ref{eqn:s_geomean}) will be
completely wrong.  Therefore, it is essential to systematically search
for all solutions, even in cases for which the event appears to be
interpretable by eye.  We do this by means of a grid search, in which
magnification maps \citep{ob05071b}
are constructed at a grid of $(s,q)$ values, and the
remaining parameters $(t_0,u_0,t_\e,\alpha,\rho)$ are seeded and then
allowed to vary in a Monte Carlo Markov chain (MCMC). The \citet{pac86}
parameters $(t_0,u_0,t_\e)$ are seeded at the 1L1S values, $\rho$
is seeded according to the prescription of \citet{gaudi02}, and
$\alpha$ is seeded at a grid of 10 values around the unit circle.

We then refine each of the local minima identified in this grid search
by seeding a new MCMC with its parameters and allowing all seven parameters
to vary.

For cases in which fitting for microlens parallax is warranted, we
initially add four parameters:
$\bpi_\e=(\pi_{\e,N},\pi_{\e,E})$ and $\bgamma=((ds/dt)/s,d\alpha/dt)$,
where \citep{gould92,gould00,gould04},
\begin{equation}
\bpi_\e = {\pi_\rel\over\theta_\e}\,{\bmu_\rel\over\mu_\rel};
\qquad 
\label{eqn:pie}
\end{equation}
and  $(ds/dt,d\alpha/dt)$ are the
first derivatives on lens orbital motion.  These vectors parameterize
the orbital effects of, respectively, Earth and the lens system, and
they can be degenerate \citep{mb09387,ob09020}.  To eliminate unphysical or
extremely rare orbits, we impose a constraint $\beta<0.8$
where $\beta$ is the absolute value of the ratio of transverse
kinetic to potential energy \citep{eb2k5,ob05071b},
\begin{equation}
\beta = \bigg|{\rm KE\over PE}\bigg|_\perp = 
{\kappa M_\odot {\rm yr}^2\over 8\pi^2}{\pi_\e\over\theta_\e}\gamma^2
\biggl({s\over \pi_\e + \pi_S/\theta_\e}\biggr)^3 ,
\label{eqn:beta}
\end{equation}
and where $\pi_S$ is the source parallax.  

For each event, we present the fit parameters of all the solutions
in table format.  In addition, each table contains the parameter
combination $t_*\equiv \rho t_\e$, which is not fit independently.

{\subsection{{KMT-2021-BLG-1391}\label{sec:anal-kb211391}}}

The overwhelming majority of the light curve (Figure~\ref{fig:1391lc})
follows a standard
1L1S profile, with
$(t_0,u_0,t_\e)\simeq (9385.29,0.012,32\,{\rm day})$.
However, within hours of the posting of the pipeline reductions, following
the alert, a short bump was noted, with center $\Delta t_{\rm anom} = -0.45\,$day
prior to $t_0$, full width $\Delta t_{\rm bump} = 0.11\,$day, and height
$\Delta I= 0.33\,$mag.  
 
{\subsubsection{{Heuristic Analysis}\label{sec:heuristic-kb211391}}}

The anomaly is at $\tau_{\rm anom} = \Delta t_{\rm anom}/t_\e = -0.014$ and hence
$u_{\rm anom} = \sqrt{\tau_{\rm anom}^2 + u_0^2} = 0.018$, i.e., at very
high magnification where planet sensitivity via central and resonant
caustics is high \citep{griest98}.  Therefore,
an important class of models that may explain this bump anomaly is that the
source has passed over a linear structure that extends from
a cusp of a central or resonant caustic (on the major image side), 
or over two very close caustics lying inside such a cusp.  In both of these 
cases, application of the
heuristic formalism of Section~\ref{sec:anal-preamble} yields,
\begin{equation}
\qquad
s_+^\dagger = 1.009.
\label{eqn:heur-kb211391}
\end{equation}

If the bump were formed by the source passing over a 
single-peak ridge, it would rise and then fall, both monotonically.
In this case, $\rho = \Delta t_{\rm bump}(\sin\alpha)/2 t_\e = 1.3\times 10^{-3}$.
In fact, the bump is flat-topped, or perhaps has a slight dimple
near its peak, which would favor the source crossing two nearly
aligned caustics just inside a cusp that are separated by of order
 the source diameter.  In this case, $\rho$ would be approximately
half as big, i.e., $\rho\sim 0.65\times 10^{-3}$.

{\subsubsection{{Static Analysis}\label{sec:static-kb211391}}}

We ultimately identified 4 solutions via a process that we describe
at the end of this subsection.  We label these as Locals 1--4.
They are illustrated in Figure~\ref{fig:1391lc}, and their parameters
are given in Table~\ref{tab:1391parms}.

In this and all tables in this paper, we generally present the median
and 68\% confidence intervals for all quantities except where otherwise
specified.  We present asymmetric intervals (in $A_{-B}^{+C}$ format)
only when the upper and lower excursions differ by more than 20\%.  Otherwise,
we symmetrize the interval (in $A\pm B$ format) to avoid clutter.  
We always present the mean and standard deviation of $\log q$ in order
to aid in understanding whether this quantity is sufficiently well
constrained to be included in mass-ratio studies.  In this case, the range of
means spans 0.12 dex (i.e., about 30\%) and their standard deviations
are substantially less, indicating that $q$ is well determined.  

We can compare these four solutions to those predicted by the 
heuristic analysis in Section~\ref{sec:heuristic-kb211391}.  Regarding $\alpha$,
all four models are nearly identical to the prediction.  From 
Table~\ref{tab:1391parms}, we see that there are two pairs of solutions:
Locals 1 and 4, have
$(s^\dagger,\Delta\ln s)_{1,4} = (1.009,0.018)$, while the 
other pair, Locals 2 and 3, have
$(s^\dagger,\Delta\ln s)_{2,3} = (1.010,0.045)$.  That is,
each pair has nearly identical $s^\dagger$ to the heuristic
prediction, but $\Delta\ln s$ is nearly 3 times larger
for the  second pair.  The biggest difference between the two
pairs is that $\rho$ is close to twice larger for the second pair.
That is, the first pair corresponds to the scenario in which the
source crosses two caustics separated by roughly the source diameter,
while the second corresponds to the caustic spacing being small
compared to the source.  See Figure~\ref{fig:1391caus}.

The first pair is preferred in the sense that it better matches 
the ``dimple'' at the midpoint of the anomaly.  However, Local 1 predicts
a slight pre-caustic depression that is not seen the data, whereas
Locals 2 and 3 do not.  Moreover, the
$\Delta\chi^2=5$ difference between Local 1 and either Local 2 or Local 3
reminds us that the advantage of Local 1 is marginal.  In any case,
both pairs have similar values of $q\sim 4\times 10^{-5}$, and all 4
solutions have even more similar values of $s$.  
The main difference between the two pairs is in the
value of $\rho$, which will propagate to factor $\sim 2$ differences
in $\theta_\e = \theta_*/\rho$ and $\mu_\rel=\theta_\e/t_\e$, which we
estimate in Section~\ref{sec:cmd-kb211391}.

We find that 1L2S solutions are excluded at $\Delta\chi^2=94.8$.

The fact that there are two pairs of degenerate solution, rather than one,
dawned on us very slowly in the course of preparing this paper.
We briefly summarize this process as it may help in the recognition
of multiple degeneracies in other events.

Originally, the grid search yielded Locals 1 and 3, as well
as a third solution that was discarded after refinement because
it had $\Delta\chi^2\sim 50$ and was a poor fit by eye.  We identified the
remaining two solutions (Locals 1 and 3, respectively) as an ``inner/outer'' 
degeneracy based partly on the morphology
shown in Figure~\ref{fig:1391caus} and partly on the fact that
the two Locals have $s_1>1$ and $s_3<1$, respectively.  Nevertheless, in the
version just prior to submission, we explicitly noted that it was puzzling that
the geometric mean $s$ for these two solutions, $\sqrt{s_1 s_3} = 0.996$, was
less than unity, even though the anomaly was clearly a major-image 
perturbation.  Our concern about this tiny discrepancy,
$|\sqrt{s_1 s_3} - s^\dagger_+|\sim 0.01$, was motivated by accumulating experience
that such ``large'' discrepancies were actually quite rare.  (It was
a similarly ``large'' discrepancy that led Gould et al. (2022, in prep)
to realize that $s^\dagger$ should be regarded as the geometric mean,
not the arithmetic mean.)

Then, in the course of final review before submission, JCY argued that
the fact that Locals 1 and 3 had substantially different $\rho$ meant
that they could not be an inner/outer (or close/wide) degeneracy.
In particular, by this time, we already had the example of 
KMT-2021-BLG-1253, which has a similar 4-fold degeneracy, comprised of
two pairs with different $\rho$, which was discovered by a completely
different route.  See Section~\ref{sec:anal-kb211253}.  We then
seeded Local 4 with the Local 1 solution 
(but $s_4 =(s^\dagger_+)^2/s_1$) and Local 2 with the Local 3 solution 
(but $s_2 =(s^\dagger_+)^2/s_3$) to find the solutions shown in 
Table~\ref{tab:1391parms}.  We discuss
some further possible implications in Section~\ref{sec:discuss}.

{\subsubsection{{Parallax Analysis}\label{sec:parallax-kb211391}}}

Formally, we find a $\Delta\chi^2=13.5$ improvement when we add the
four parameters $\bpi_\e$ and $\bgamma$.  However, there is very strong
evidence that this ``measurement'' is due to low-level systematics
rather than real physical effects.  First, even if the errors were truly
Gaussian, the probability of such a $\Delta\chi^2$ for four degrees of
freedom (dof) would be $p=(1 + \Delta\chi^2/2)\exp(-\Delta\chi^2/2) =1\%$.
In itself, this would be regarded as strong, but not overwhelming evidence
for the measurement.  Second, however, the plots of cumulative 
$\Delta\chi^2$ with time for the three observatories (not shown) are not 
consistent  with each other, demonstrating
that the assumption of Gaussian statistics is 
too strong.  Third, the derived parallax values are very high, $\pi_\e\sim 4$.
In Section~\ref{sec:cmd-kb211391}, we will show that either 
$\theta_\e\sim 0.5\,\mas$ or $\theta_\e\sim 0.25\,\mas$.  This parallax
determination would then imply 
$(M,D_L)\sim  (16\,M_{\rm jupiter},0.5\,\kpc)$ or
$(M,D_L)\sim  (8\,M_{\rm jupiter},1\,\kpc)$.  The prior probability
of such nearby lenses is extremely small and would by itself
counterbalance the above $p=1\%$, even if Gaussian statistics applied.
Further, the prior probability for such low-mass free-floating-planet-like
objects (with Earth-mass moons!) is also very low.
Finally, the parallax signal comes from the wings of the event, where
(because of the faint source, $I\sim 22$) the statistical errors are much
larger than the difference fluxes.  This is a regime where low-level
red noise can easily give rise to spurious signals.  We therefore
adopt the static solutions.


{\subsection{{KMT-2021-BLG-1253}\label{sec:anal-kb211253}}}

The majority of the light curve (Figure~\ref{fig:1253lc}) follows a standard
1L1S profile, with
$(t_0,u_0,t_\e)\simeq (9374.41,0.006,9\,{\rm day})$.
However, just after peak, there is
a short bump, with center $\Delta t_{\rm anom} = +0.035\,$day
after $t_0$, full width $\Delta t_{\rm bump} = 0.057\,$day, and height
$\Delta I= 0.26\,$mag.  

{\subsubsection{{Heuristic Analysis}\label{sec:heuristic-kb211253}}}

The anomaly is at $\tau_{\rm anom} = \Delta t_{\rm anom}/t_\e = 0.0039$ and hence
$u_{\rm anom} = \sqrt{\tau_{\rm anom}^2 + u_0^2} = 0.0071$, i.e., at very
high magnification. Thus, similarly to KMT-2021-BLG-1391, we expect that
this bump anomaly may be explained by the
source passing over a linear structure that extends from
a cusp of a central or resonant caustic on the major image side, 
or over two very close caustics lying inside such a cusp.  Applying
the heuristic formalism of Section~\ref{sec:anal-preamble}, we obtain,
\begin{equation}
\alpha = 57^\circ; \qquad 
s_+^\dagger = 1.004 .
\label{eqn:heur-kb211253}
\end{equation}

As with KMT-2021-BLG-1391, if the source crosses a cusp, then
we expect $\rho \sim \Delta t_{\rm bump}\sin\alpha/2 t_\e = 2.6\times 10^{-3}$,
while it should be half this value, $\rho\sim 1.3\times 10^{-3}$, if the
source crosses a pair of closely spaced caustics.

{\subsubsection{{Static Analysis}\label{sec:static-kb211253}}}

The initial grid search yields four solutions.  After refining
these, we find that one of these solutions is disfavored by $\Delta\chi^2=63$
and also appears by eye to be a poor fit to the data.  We therefore do not 
further consider it.  We notice that one pair of solutions
(called ``Local 1'' and ``Local 2'') are very nearby in $(s,q)$, which raises
the concern that they might not be truly distinct.  To address this, we
run a ``hot chain'' (i.e., artificially increase the error bars) so that the 
chain will move easily between minima.  We then find that these two solutions 
have a strong ($\Delta\chi^2\sim 15$) barrier between them, so they are indeed
distinct.  Note, however, that if this event had been in a 
$\Gamma=1\,{\rm hr}^{-1}$ cadence field, the barrier would have been
four times smaller, which would have implied semi-merged minima.
After applying a similar hot chain to Local 3, we find a 
fourth solution, which is a very localized and fairly weak minimum.
This emphasizes the importance of running such hot
chains, particularly on solutions that have resonant and near-resonant
topologies, for which there can be multiple caustic/trajectory geometries
that produce similar light curves.  Note that if the original grid search
had not identified Local 1, then only hot chains that were carried out as a 
matter of ``due diligence'' would have uncovered the extra solutions.

From Table~\ref{tab:1253parms}, 
we see that the pair, Locals 1 and 4, have
$(s^\dagger,\Delta\ln s)_{1,4} = (1.004,0.068)$, while the 
pair, Locals 2 and 3, have
$(s^\dagger,\Delta\ln s)_{2,3} = (1.002,0.138)$.  That is, both have the 
$s^\dagger\simeq s^\dagger_+$ that was predicted by the heuristic analysis,
but with $\Delta\ln s$ values that differ by a factor 2 from each other.
The former set of solutions have $\rho\sim 1.3\times 10^{-3}$, in good
agreement with the prediction for the source crossing a close pair of caustics,
while the latter have much larger $\rho\sim 2.0\times 10^{-3}$, in qualitative
agreement with the prediction for a cusp-crossing geometry.
See Figure~\ref{fig:1253cau}.
All four solutions have trajectory angles $\alpha$ in good agreement with
the heuristic analysis.

All four solutions have $q\simeq 2.3\times 10^{-4}$ (within errors), and
they also have qualitatively similar $s$.  As was the case for 
KMT-2021-BLG-1391, the main difference is in the values of $\rho$, which
vary by about a factor 1.5 among the solutions and will propagate into
similar differences in $\theta_\e$ and $\mu_\rel$.

We find that 1L2S solutions are excluded at $\Delta\chi^2=69.9$.

{\subsubsection{{Parallax Analysis}\label{sec:parallax-kb211253}}}

Given the event's very short timescale ($t_\e\sim 9\,$days) and very faint
source $I_s\sim 23.2$, the difference fluxes are below the statistical
errors even three days from the peak.  The annual parallax signal
is negligible on such short timescales, so we do not attempt a
parallax measurement, and we therefore adopt the static solutions.

{\subsection{{KMT-2021-BLG-1372}\label{sec:anal-kb211372}}}

The light curve (Figure~\ref{fig:1372lc}) mostly follows a 1L1S profile 
$(t_0,u_0,t_\e)\simeq (9388.8,0.075,71\,{\rm days})$,
with the dip of about 0.5 mag, lasting about $\Delta t_{\rm dip}=1.5\,$days, 
centered at $t_{\rm dip} = 9387.70$.  

{\subsubsection{{Heuristic Analysis}\label{sec:heuristic-kb211372}}}

Noting that $\tau_{\rm anom} = (t_{\rm anom} - t_0)/t_\e = -0.015$,
the heuristic formalism of Section~\ref{sec:anal-preamble} implies,
\begin{equation}
\alpha = 282^\circ; \qquad 
s_-^\dagger = 0.962;
\qquad q \sim 3.4\times 10^{-4}.
\label{eqn:heur-kb211372}
\end{equation}

{\subsubsection{{Static Analysis}\label{sec:static-kb211372}}}

The grid search returns two solutions, which after refinement,
yield $(s^\dagger,\Delta\ln s)=(0.963,0.048)$ and $\alpha=281^\circ$,
in excellent agreement with the heuristic prediction, while the values
of $q\simeq 4.3\times 10^{-4}$ are in good agreement.  
See Table~\ref{tab:1372parms}.

We note that there is an extraordinarily bright 
($G=9.0$, $J=6.1$) star about $23^{\prime\prime}$
from the event whose diffraction spikes create some difficulties for
the photometry.  We find that, for KMTC, pyDIA handles these difficulties
better than pySIS.  Hence, we use pyDIA, and this is the system that we
quote in Table~\ref{tab:1372parms}.

{\subsubsection{{Parallax Analysis}\label{sec:parallax-kb211372}}}

Motivated by the long timescale ($t_\e\sim 70\,$days), we attempt
to measure the microlens parallax and, as usual begin by including
orbital motion in the fit. However, we find that neither is usefully
constrained in the joint fit and even when we suppress the orbital
degrees of freedom, the parallax is still not usefully constrained.
As parallax is mostly constrained from the wings of the light curve
where the source is only slightly magnified relative to its baseline
value $I_S\sim 22$, the faintness of the source is likely the origin
of the difficulty of making the parallax measurement, possibly compounded
by the diffraction spikes.  In any case,
we adopt the static analysis of Section~\ref{sec:static-kb211372}.

{\subsection{{KMT-2021-BLG-0748}\label{sec:anal-kb210748}}}

The light curve follows a low-magnification 1L1S profile,
$(t_0,u_0,t_\e) = (9344.6,0.39,41\,{\rm days})$, except for a
short bump represented by three points centered at $t_{\rm bump}\simeq 9346.65$.
The three points span only 3.6 hours, but the bump is confined to
a total of $\Delta t_{\rm bump}<8.6\,$hours by two flanking points.  
Thus it is reasonably
well characterized despite the relatively low nominal cadence of the
the field, $\Gamma = 0.4\,{\rm hr}^{-1}$.

{\subsubsection{{Heuristic Analysis}\label{sec:heuristic-kb210748}}}

The heuristic formalism yields
\begin{equation}
\alpha = 83^\circ; 
\qquad s_+^\dagger = 1.216.
\label{eqn:heur-kb210748}
\end{equation}
We also obtain 
$\rho \simeq t_{\rm bump}\sin\alpha/2 t_\e \la 4.3\times 10^{-3}$.

{\subsubsection{{Static Analysis}\label{sec:static-kb210748}}}

The grid search appears (see below) to return two solutions,
whose parameters, after refinement, are shown as Locals 1 and 2 in 
Table~\ref{tab:0748parms}.
These have $(s^\dagger,\Delta\ln s)=(1.228,0.031)$ and $\alpha\simeq 83^\circ$,
in good agreement with Equation~(\ref{eqn:heur-kb210748}).

However, as a matter of due diligence, we run a hot chain at these minima
and identify two ``satellite solutions'', whose parameters are given in 
Table~\ref{tab:0748parms} as Locals 3 and 4.  In retrospect, we noted that these
solutions actually do appear in the grid search (not shown), but because they
have substantially worse fits ($\Delta\chi^2\sim 10$), they tended to appear
as extended halos of the two best solutions in the grid display rather
than as distinct minima.  This again emphasizes the importance of
investigating hot chains around identified solutions.

These two additional solutions have similar $\alpha\simeq 84^\circ$, but
quite different $s^\dagger$, i.e., $(s^\dagger,\Delta\ln s)=(1.446,0.0006)$.
More significantly, their mass ratios, $q$ are different by factors of several
and their self crossing times $t_*\simeq 3.4\,$hr are more than twice 
those of the best solutions, $t_*\simeq 1.5\,$hr.  

The latter difference is the key to understanding the relation between the
two sets of solutions.  
In the pair of best solutions, the bump is due to the source crossing
two nearly parallel caustics, just inside the inner (or outer) on-axis cusp 
of the planetary caustic.  Hence, $t_*$ is about 1/4 of the full duration
of the bump.  In the satellite solutions, the source crosses the cusp, so
$t_*$ is about 1/2 the full duration of the bump.  To accommodate these
different morphologies, both $q$ and $t_\e$ (and so $u_0$ and $I_S$) 
are also affected.

Our basic orientation toward these satellite solutions is that they
should be considered as $3\,\sigma$ variations on the best solutions.
That is, even though they are topologically isolated in the 7-dimensional
space of solutions, they represent normal variations on the best solutions.
Therefore, we adopt the parameters of the best solutions for this planet.

Nevertheless, we will show in Section~\ref{sec:phys-kb210748} how future AO
observations can definitively resolve the issue (presumably by rejecting
the satellite solutions).

We find that 1L2S solutions are excluded at $\Delta\chi^2=273$.

{\subsubsection{{Parallax Analysis}\label{sec:parallax-kb210748}}}

In view of the low amplitude and faint source of this event, we
do not attempt a parallax analysis.  We therefore adopt the static-model
results presented in Section~\ref{sec:static-kb210748}.

\section{{Source Properties}
\label{sec:cmd}}

Our primary aim in analyzing the color-magnitude diagram (CMD) of each
event is to measure $\theta_*$ and so determine
\begin{equation}
\theta_\e = {\theta_*\over\rho};\
\qquad
\mu_\rel = {\theta_\e\over t_\e}.
\label{eqn:thetae_murel}
\end{equation}
We follow the method of \citet{ob03262} by first finding the offset
of the source from the red clump
\begin{equation}
\Delta[(V-I),I] = [(V-I),I]_S - [(V-I),I]_{\rm cl},
\label{eqn:deltacmd}
\end{equation}
where we adopt  $(V-I)_{\rm cl,0}=1.06$ from \citet{bensby13}, and we evaluate
$I_{\rm cl,0}$ from Table~1 of \citet{nataf13}, based on the Galactic
longitude of the event.  This yields the  dereddened color and magnitude
of the source
\begin{equation}
[(V-I),I]_{S,0} = [(V-I),I]_{\rm cl,0} + \Delta[(V-I),I].
\label{eqn:cmd0}
\end{equation}
We transform from $V/I$ to $V/K$ using the $VIK$ color-color 
relations of \citet{bb88}, and then apply the color/surface-brightness relations
of \citet{kervella04} to obtain $\theta_*$.  After propagating the
measurement errors, we add 5\% to the error in quadrature to take account
of systematic errors due to the method as a whole.

To obtain $[(V-I),I]_S$, we generally begin with pyDIA reductions \citep{pydia},
which put the light curve and field-star photometry on the same system.
We determine $(V-I)_S$ by regression of the $V$-band data 
on the $I$-band data, and we determine $I_S$ by regression of the $I$-band
data on the best-fit model.  When there is calibrated OGLE-III 
field-star photometry \citep{oiiicat}, we transform $[(V-I),I]_s$ to
this system.  Otherwise we work in the instrumental KMT pyDIA system,
which typically has offsets of $\la 0.2\,$mag from the standard system.
The CMDs are shown in 
Figure~\ref{fig:allcmd}.
When OGLE-III photometry is available, we choose a radius that balances
the competing demands of reducing differential extinction and
having sufficient density in the clump to measure its centroid.

In only one case (KMT-21-BLG-1391) are we able to make a good measurement
of the source color from regression.  In the remaining three cases,
we estimate this color from the offset of the $I$-band source magnitude relative
to the clump.  This offset is usually determined (as indicated above)
from the clump position found in the $[(V-I),I]$ color-magnitude diagram.
However, in two cases (KMT-2021-BLG-1253 and KMT-2021-BLG-1372), we cannot
confidently measure the clump position in these optical bands.  Therefore,
for these cases, we determine $I_{\rm cl}$ from a $[(I-K),I]$ CMD.

The logical chains leading to the $\theta_*$ measurements are 
encapsulated in Table~\ref{tab:cmd}.   In all cases, the source
flux is that of the best solution.  Under the assumption of fixed
source color, $\theta_*$ scales as $10^{-\Delta I_S/5}$ for the other solutions,
where $\Delta I_S$ is the difference in source magnitudes, as given
in the Tables of Section~\ref{sec:anal}.
The inferred values (or limits
upon) $\theta_\e$, and $\mu_\rel$ are given in the 
individual events subsections below, where we also discuss other
issues related to the CMDs of each event.

{\subsection{{KMT-2021-BLG-1391}}
\label{sec:cmd-kb211391}}

The values of the source flux and resulting $\theta_*$ that are given
in Table~\ref{tab:cmd} are for Local 1, which is
preferred by $\Delta\chi^2=5$.  We note that for Local 4, $I_S$ is
fainter by 0.005 mag, implying that $\theta_*$ is smaller by 0.2\%.
This change is far too small to make any difference in the present case.
In general, however, the CMD analysis in Table~\ref{tab:cmd} will always
give values for the best-fit solution, and $\theta_*$ for other solutions
can be inferred by $\theta_*\propto 10^{-\Delta I/5}$.
Then, using the values in Table~\ref{tab:1391parms}, we find that
$\theta_\e=\theta_*/\rho$ and $\mu_\rel=\theta_\e/t_\e$ are given by,
\begin{equation}
\theta_\e =0.456\pm 0.054\,\mas,
\qquad
\mu_\rel =5.26\pm0.62 \,\masyr,
\qquad
({\rm Local\ 1}),
\label{eqn:1391L1parms}
\end{equation}
\begin{equation}
\theta_\e =0.226\pm 0.025\,\mas,
\qquad
\mu_\rel = 2.61\pm 0.29\,\masyr,
\qquad
({\rm Local\ 2}).
\label{eqn:1391L2parms}
\end{equation}
\begin{equation}
\theta_\e =0.227\pm 0.025\,\mas,
\qquad
\mu_\rel = 2.61\pm 0.29\,\masyr,
\qquad
({\rm Local\ 3}),
\label{eqn:1391L3parms}
\end{equation}
\begin{equation}
\theta_\e =0.426\pm 0.051\,\mas,
\qquad
\mu_\rel =4.90\pm0.58 \,\masyr,
\qquad
({\rm Local\ 4}).
\label{eqn:1391L4parms}
\end{equation}

We align the position of the source when it was well magnified
(so, easily measured) to baseline images taken at the 
3.6m Canada-France-Hawaii Telescope (CFHT), in which the 
$I_{\rm catalog}\sim 19.2$ 
catalog object appears isolated in $0.5^{\prime\prime}$ seeing.  The source
position is offset from this object by $0.7^{\prime\prime}$,
so it clearly is not associated with the event.  There is no obvious
object at the position of the source, and there is clearly less flux
at this position than from a neighboring star with measured
magnitude $I_{\rm neighbor}=21.04\pm 0.14$, while the faintest star detected
in the image is $I_{\rm faintest} =21.91$.  Hence we can place a firm
lower limit on the magnitude of the baseline object, $I_{\rm base}>21$, and 
therefore on the blend $I_b> 21.6$.  This is a significant 
constraint: for a lens at the distance of the Galactic bar, 
this would imply $M_I > 6.1$, which effectively rules out lenses
with $M> 0.7\,M_\odot$.  We will incorporate this flux constraint directly
into the Bayesian analysis in Section~\ref{sec:phys-kb211391}.

While we cannot distinguish between the two solutions because
$\Delta\chi^2$ is only 5, they can ultimately be distinguished based
on AO follow-up observations on, e.g., 30m class telescopes, because the
two solutions predict very different proper motions.  
See Section~\ref{sec:phys-kb211391}.

{\subsection{{KMT-2021-BLG-1253}}
\label{sec:cmd-kb211253}}

As discussed above, heavy extinction prevents us from measuring 
$(V-I)_S$ from the light curve.  Moreover, while the clump can
be discerned on the pyDIA CMD (not shown), it is truncated by
the $V$-band detection limit, so its centroid cannot be reliably
measured.  We therefore derive $I_{\rm cl}$ by combining $I$-band
data from OGLE-III and $K$-band data from the VVV survey \citep{vvvcat}
to construct an $[(I-K),K]$ CMD.  See Figure~\ref{fig:allcmd}.  After 
transforming the pyDIA $I_S$ measurement
to the OGLE-III system, we find that the source
lies $\Delta I=4.67$ magnitudes below the clump (see Table~\ref{tab:cmd}).
Making use of the {\it Hubble Space Telescope (HST)} CMD of Baade's window
\citep{holtzman98},
we estimate $(V-I)_0= 0.80\pm 0.10$.  In Figure~\ref{fig:allcmd},
we display the corresponding $(I-K)_s$, which we derive from \citet{bb88}
and the observed $(I-K)_{\rm cl} =5.33$ color of the clump.  We then
proceed as usual.

The values of the source flux and resulting $\theta_*$ that are given
in Table~\ref{tab:cmd} are for Local 1, which is
preferred by just $\Delta\chi^2<1.6$ over Local 3 but has
a $t_* \equiv\rho t_\e$ that is a factor 0.6 smaller.
As in the case of KMT-2021-BLG-1391, the evaluations for the
four solutions below reflect the differences in $I_S$ as well as in $\rho$:
\begin{equation}
\theta_\e = 0.399\pm 0.062\,\mas, \qquad
\mu_\rel = 15.2\pm 2.4 \,\masyr \qquad {\rm (Local\ 1)},
\label{eqn:1253L1parms}
\end{equation}
\begin{equation}
\theta_\e = 0.250\pm 0.036\,\mas, \qquad
\mu_\rel = 9.0\pm 1.3 \,\masyr \qquad {\rm (Local\ 2)},
\label{eqn:1253L2parms}
\end{equation}
\begin{equation}
\theta_\e = 0.244\pm 0.034\,\mas, \qquad
\mu_\rel = 9.4\pm 1.3 \,\masyr \qquad {\rm (Local\ 3)},
\label{eqn:1253L3parms}
\end{equation}
\begin{equation}
\theta_\e = 0.405\pm 0.063\,\mas, \qquad
\mu_\rel = 15.5\pm 2.4 \,\masyr, \qquad {\rm (Local\ 4)}.
\label{eqn:1253L4parms}
\end{equation}

The two pairs of solutions predict very different lens-source relative
proper motions, $\mu_\rel$, and they can therefore be distinguished
by future AO follow-up observations after elapsed time $\Delta t$,
when the lens and source have separated by $\Delta\theta=\mu_\rel\Delta t $.  
However, as we discuss in Section~\ref{sec:phys-kb211253}, this will be of
less direct interest than in the case of KMT-2021-BLG-1391
because all four solutions have essentially the same
mass ratio $q\simeq 2.3\times 10^{-4}$ and normalized projected separation
$s\simeq 1$.

As with KMT-2021-BLG-1391, we align the position of the source 
when it was well magnified
to archival images taken at the CFHT.  Similarly to that case, the
$I_{\rm catalog}\sim 19.4$ 
catalog object appears relatively isolated in $0.5^{\prime\prime}$ seeing.  
However, in this case, the source
position is offset from the catalog star by $0.49^{\prime\prime}$,
meaning that the catalog star somewhat overlaps the source position.
The catalog star is just 0.8 mag fainter than the clump, i.e., $>3$ mag
brighter than a solar mass main-sequence star in the bulge.  Hence,
contamination by the catalog star prevents us from placing any
useful limits on light from the lens, even in these excellent CFHT
images.

{\subsection{{KMT-2021-BLG-1372}}
\label{sec:cmd-kb211372}}

Due to high extinction $A_I\sim 3.3$ and the relatively faint peak
of the event, $I_{\rm peak}\sim 18.4$, we expect roughly $(V-I)_s\sim 4.1$
and thus $V_{\rm peak}\sim 22.5$.  Therefore, we do not expect to be able
to measure $(V-I)_S$ and, unfortunately, this proves to be the case.
Moreover, the expected position of the clump at $V_{\rm cl}\sim 22$
is too faint to be reliably defined on the $[(V-I),I]$ CMD.
Hence, as for KMT-2021-BLG-1253, we carry out the color-magnitude
analysis on an $I$ vs.\ $(I-K)$ CMD.  As in that case, we extract $K$-band
field star photometry from VVV \citep{vvvcat}, and we infer the source
color from its $I$-band offset relative to the clump.  In contrast to
the case of KMT-2021-BLG-1253, however, there is no OGLE-III photometry
for this field, and we therefore cannot calibrate our KMTC pyDIA photometry.
Hence, we report it directly.  Nevertheless, this does not pose any
great difficulties because, first, our estimates of $\theta_*$ depend
only on relative photometry, and, second, for the cases that we can perform
such calibrations, the typical offset is about $-0.2$ magnitudes.
In particular, for the four other events for which we carried out detailed
CMD analyses in the preparation of this paper\footnote{The three other planetary
events, plus KMT-2021-BLG-0278, which was eliminated at a late stage 
(see Section~\ref{sec:intro}).}, we find that
the mean and standard deviation of this offset are
$I_{\rm OGLE-III} - I_{\rm pyDIA,KMTC} = -0.18\pm 0.05$.

Because the source color is not directly measured,
we use the offset from the clump, $\Delta I= 3.93$,
and the \citet{holtzman98} {\it HST} CMD to estimate 
$(V-I)_{S,0}= 0.75\pm 0.10$.

We obtain only an upper limit on $\rho$, which is better expressed
as $t_* < 0.3\,$day at $2.5\,\sigma$.  This arises from the fact that
for larger $t_*$, the pre-dip ``bump'' at about 
HJD$^\prime={\rm HJD}-2450000 = 9387$ would be washed out.
This leads to a limit on the proper motion,
$\mu_\rel>\mu_{\rm lim}= 0.84\,\masyr$, which is relatively unconstraining.  
That is, a fraction 
$p < (\mu_{\rm lim}/\sigma_\mu)^3/6\sqrt{\pi} \rightarrow 2\times 10^{-3}$ of all 
microlensing events have such low proper motions (Gould et al. 2022, in prep),
where we have approximated the bulge proper motion distribution as having
a dispersion of $\sigma_\mu=2.9\,\masyr$ in each direction.

We align 5 archival CFHT images, taken in $0.45^{\prime\prime}$ to
$0.5^{\prime\prime}$ seeing to the KMT field, in which the source position
is very well determined from high-magnification difference images.
We find that the source lies $0.71^{\prime\prime} \pm 0.03^{\prime\prime}$
due west of the $I=21.3$ (KMT system) catalog star, whose monitoring
permitted detection of the event.  The source position shows no detectable
flux, either from the wings of the catalog star or from the source or lens,
from which we put a lower limit on the baseline magnitude 
$I_{\rm base,kmt} > 21.8$, i.e., similar to $I_S$ (see Table~\ref{tab:cmd}).
We adopt a conservative limit $I_{L,\rm kmt} \ga 22.5$, which
as indicated above (based on the calibrations of other fields), 
corresponds to $I_{L,\rm calib} \ga 22.3$.

{\subsection{{KMT-2021-BLG-0748}}
\label{sec:cmd-kb210748}}

As in the case of KMT-2021-BLG-0748, the $[(V-I),I]$ CMD is well characterized,
but the source color $(V-I)_S$ cannot be directly measured from regression
of the $V$- and $I$-band light curves.  The reason for this is similar:
The $I$-band flux variation at peak corresponds to a difference magnitude
only $\delta I_{\rm peak} = 20.3$.  Taking account of the extinction and
typical source colors, we expect $(V-I)_S \sim 2.0$, implying
$\delta V_{\rm peak} \sim 22.3$, which is too faint for a reliable measurement.
Because the source lies $\Delta I=4.9$ below the clump, we estimate
its color using the \citet{holtzman98} {\it HST} CMD to be
$(V-I)_{S,0} = 0.83\pm 0.10$.

The two principal solutions have very similar inferred 
$\theta_\e$ and $\mu_\rel$ (differences that are an order of
magnitude smaller than their errors), for which we therefore present 
the average:
\begin{equation}
\theta_\e = 0.32\pm 0.06\,\mas, \qquad
\mu_\rel = 2.9\pm 0.6 \,\masyr \qquad {\rm (Principal\ Solutions\ [L1+L2])}.
\label{eqn:0748principal}
\end{equation}
Although we do not consider the satellite solutions to be truly 
independent and believe that their $\chi^2$ values should be
interpreted at face value, we nonetheless present their corresponding
implications in order to guide the interpretation of future AO observations,
as discussed in Section~\ref{sec:phys-kb210748}:
\begin{equation}
\theta_\e = 0.16\pm 0.03\,\mas, \qquad
\mu_\rel = 2.1\pm 0.3 \,\masyr \qquad {\rm (Satellite\ Solutions\ [L3+L4])}.
\label{eqn:0748satellite}
\end{equation}

\section{{Physical Parameters}
\label{sec:phys}}

None of the four planets have sufficient information to precisely
specify the host mass and distance.  Moreover, several have multiple
solutions with significantly different mass ratios $q$ and/or different 
Einstein radii $\theta_\e$.  For any given solution, we can incorporate
Galactic-model priors into standard Bayesian techniques to obtain estimates
of the host mass $M_{\rm host}$ and distance $D_L$, 
as well as the planet mass $M_{\rm planet}$ 
and planet-host projected separation $a_\perp$.
See \citet{ob180567} for a description of the Galactic model and Bayesian
techniques.  However, in most cases we still have to decide how to
combine these separate estimates into a single ``quotable result''.
Moreover, in several cases, we also discuss how the nature of the planetary
systems can ultimately be resolved by future AO observations.  Hence,
we discuss each event separately below.

{\subsection{{KMT-2021-BLG-1391}}
\label{sec:phys-kb211391}}

There are two pairs of solutions whose only major differences are
$\rho_{2,3}/\rho_{1,4} = 2.0\pm 0.2$ and $q_{2,3}/q_{1,4} = 1.28 \pm 0.17$.
In order to be relevant to both present and future (post AO follow-up)
studies, we report five Bayesian results in Table~\ref{tab:physall}, 
one for each Local 
and one weighted average.  The weighting has two factors.  One factor
comes from the difference of the $\chi^2$ minima of the different Locals.
For example, it disfavors Local 2 relative to Local 1 by
$\exp(-\Delta\chi^2/2)=0.08$.  The other
comes from the fraction of Bayesian trials that are consistent with the
measured parameters $t_\e$ and $\theta_\e$, and with the flux constraint.  
These are (by chance, see below) nearly identical.

As discussed in Section~\ref{sec:cmd-kb211391}, 
there is a strict upper limit on
lens flux, $I_L > 21.6$.  We incorporate this limit 
by assuming that the $A_I=1.19$
magnitudes of extinction along the line of sight (see Table~\ref{tab:cmd}),
are generated by a dust profile with an $h=120\,$pc scale height.
The flux constraint eliminates more statistical weight from Locals 1 and 4
simulations because they have higher $\theta_\e$ and are therefore more
populated with the $M\ga 0.7\,M_\odot$ lenses that are eliminated by
this constraint.  This roughly cancels the effect of 
the higher $\theta_\e$ (and $\mu_\rel$), which would otherwise
favor these Locals 
by increasing the available phase space and the cross section.

For present-day studies, we recommend the combined solution for 
estimates of the physical parameters.  The four individual solutions
may be useful for comparison with future AO observations that separately
resolve the source and lens. As mentioned above, the two pairs of solutions
predict very different values of $\mu_\rel$, which will be decisively
measured by future AO observations.  In addition, the corresponding $\theta_\e$
(also very different between the two pairs) and lens flux (and
possibly color) must yield a self-consistent lens mass and distance.

We note that if the favored small-$\rho$ solutions are correct, then in 2030
(plausible first AO light for MICADO), the lens and source will
be separated by $\Delta\theta\sim 47\pm 6\,\mas$, while for the
disfavored large-$\rho$ solutions, $\Delta\theta\sim 23\pm 3\,\mas$.  In 
both cases, this is well above the $H$-band FWHM$\sim 11\,\mas$
and more so for the $J$ band.
Such imaging will be required to accurately determine the host mass.   
The combination of this mass measurement, together with distinguishing 
between the two pairs of solutions, 
will yield either an 8\% or 13\% error in the planet
mass, depending on which solution proves to be correct.

{\subsection{{KMT-2021-BLG-1253}}
\label{sec:phys-kb211253}}

The situation for KMT-2021-BLG-1253 is broadly similar to that of
KMT-2021-BLG-1391: there are four solutions whose parameters are similar,
with the exception of $\rho$.  
Also note that the $\chi^2$ values differ only modestly
among the four solutions.  However, in this case, 
there is no useful limit on lens flux.

We proceed in essentially the same way.  Table~\ref{tab:physall} shows the 
results for the four separate solutions individually, as well as the combined
solution.  We note that Locals 1 and 4 are significantly disfavored
by their relatively high proper motions combined with relatively small
Einstein radii.  The former are easily generated for nearby lenses,
$D_L\la 2\,\kpc$, but in this case the corresponding $\theta_\e$ would
imply $M\la 0.05\,M_\odot$ hosts, which are relatively rare.  On the
other hand, for bulge-bulge lensing (which often generates such $\theta_\e$),
such high $\mu_\rel$ are relatively rare.  

As in the case of KMT-2021-BLG-1391, future AO observations will decisively
distinguish between the high- and low-$\mu_\rel$ solutions.  However,
this is of less direct interest in the current case because these two
classes of solutions do not differ significantly in $q$.  Hence, the
principal interest will be (as usual) to measure the lens mass from
a combination of the lens flux and the more precisely determined $\theta_\e$
(from the more precisely measured $\mu_\rel$, together with $t_\e$).

Note that while the catalog star is relatively bright ($I\sim 19.4$,
likely $K\sim 14$), it is displaced from the lens-source system
by $0.5^{\prime\prime}$.  Thus, it is unlikely to pose any challenges
to detecting the lens with 30m class telescopes, 
even if the lens proves to be a faint bulge 
dwarf, $M_K\sim 9$ ($K\sim 24$), and so has a contrast ratio of 10,000.  
However, this issue would need to be carefully considered for current
10m class telescopes.

{\subsection{{KMT-2021-BLG-1372}}
\label{sec:phys-kb211372}}

In this case, the two degenerate solutions are essentially identical,
except that they differ in $s$ by about 12\%.  Hence, the physical
parameter estimates in Table~\ref{tab:physall} are also nearly identical.
As mentioned in Section~\ref{sec:cmd-kb211372}, we impose two limits
(in addition to the $t_\e$ measurement).  First, we impose $t_*< 0.3\,$day,
which we argued is hardly constraining.  Second, we impose $I_L>22.3$.
This also has very little practical effect because of the high extinction,
$A_I = 3.93$ (see Table~\ref{tab:cmd}).  As a sanity check, we explore
the effect of imposing a stronger limit $I_L>23$, which might plausibly
be inferred from the upper limit on the baseline flux.  However, this
results in a reduction in the Bayesian mass estimates of only 10\%,
i.e., 5 times smaller than the uncertainty, which again illustrates
that the lens flux constraint is weak.  We refrain from imposing this
stronger constraint because it is less secure and would have very little
effect.

The long timescale of this event ($t_\e\sim 70\,$days) most likely
implies a low proper motion.  In the Bayesian analysis, we find
$\mu_\rel= 2.75\pm 1.25\,\masyr$.  Because the lens is likely to be 
fainter than the source in $I$ and also redder than the source,
and because of high-extinction, the AO followup observations should
probably be done in $H$ or $K$, for which the FWHM, even on the 39m 
European ELT (EELT),
are respectively, 10 mas and 14 mas.  Thus, it is possible, but far
from guaranteed that the lens can be separately resolved in, e.g., 2030.

{\subsection{{KMT-2021-BLG-0748}}
\label{sec:phys-kb210748}}

In Table~\ref{tab:physall}, we show the Bayesian estimates of the two principal
solutions and also their weighted average.  As discussed in 
Section~\ref{sec:cmd-kb210748}, we do not consider the satellite solutions
to be independent, and we therefore do not separately display their
physical-parameter estimates.  We note, however, that if they were
included in the weighted average, their contribution would be negligible,
being downgraded by $\exp(-\Delta\chi^2/2)\sim 0.004$, and by a further
factor from the Galactic model, due to their lower proper motions.

Given the relatively low proper motion, $\mu_\rel\sim 2.9\,\masyr$, of 
even the principal solutions, it will not be possible to separately
resolve the lens and source until first AO light on 30m class telescopes.
At that point, the principal interest will, as usual, be to measure
the lens flux, which when combined with the already measured $\theta_\e$,
will yield the lens mass and distance.  However, the AO-based $\mu_\rel$
measurement will also improve the determination of 
$\theta_\e = \mu_\rel t_\e$ because $t_\e$ is much better measured than
$\theta_*$ (due in part to our inability to directly measure the source color).

In addition, the AO measurement can also more decisively exclude the
satellite solutions.  While the AO measurement of $\mu_\rel$ can only
marginally distinguish between the predictions of 
Equations~(\ref{eqn:0748principal}) and (\ref{eqn:0748satellite}),
the source-flux predictions of the two sets of solutions differ
by more than a factor 2.  While this comparison will be somewhat
blurred by the fact that the AO flux measurement will be in an infrared
band (whereas the microlensing model estimates are in the $I$ band),
the uncertainty induced by the lack of a color measurement will be
$<0.2$ mag, i.e., much smaller than the factor 2 difference in flux predictions.

\section{{Discussion}
\label{sec:discuss}}

We have presented solutions and analysis for 4 microlensing planets
that were discovered from KMTNet data during the 2021 season, as the
events were evolving.
In contrast to most previous papers about microlensing planets, our goal is not
to highlight the features of the individual discoveries but rather
to expedite the publication of all 2021 KMTNet planets in order to enable
future statistical studies.  There will be several dozen such planets.
Hence, this paper constitutes, in the first place, a ``down payment'' on
that effort.  In addition to the dozens of planets that have been identified
by eye from AlertFinder \citep{alertfinder} events, there may be additional
planets discovered after the EventFinder \citep{eventfinder} search has been
completed, and more yet after AnomalyFinder \citep{ob191053} has been
run on all events.  Hence, we consider it prudent to begin ``mass production''
of planets quickly.

Although the choice of events was simply to rank order those identified
by YHR according to their initial mass-ratio estimates, there are some
interesting features of the sample as a whole.  First, two of the events
(KMT-2021-BLG-1391 and KMT-2021-BLG-1253) suffer from a discrete 
degeneracy in $\rho$ in which a short bump is explained either by a 
larger-$\rho$ source that straddles two closely spaced caustics or by a
smaller-$\rho$ source that successively passes two caustics that are separated
by roughly one source diameter.  The case of KMT-2021-BLG-1253 is particularly
striking because (in contrast to KMT-2021-BLG-1391) there are no identifiable
features that could have qualitatively distinguished between the two models 
if there had been more data.  To our knowledge this is the first report
of this degeneracy.  In any case, it is certainly not so common that
we would have expected 2 examples in a random sample of 4 events.  This
``high rate'' may be just due to chance, but it could also indicate that
alternate solutions have been missed in past events.  In both cases,
the different values of $\rho$ lead to very different proper motions.
Hence the degeneracy can be broken by future AO observations that
separately resolve the lens and source.

This paper is the first to adopt the orientation that essentially all 
of the planets being reported now can yield mass (rather than mass ratio)
measurements in less than a decade and therefore to provide systematic
support and guidance for such observations.  The first case of source-lens
resolution was made more than 20 years ago \citep{alcock01}, and there has
been a steady stream of such measurements for microlensing planets over the
past 7 years \citep{ob05169ben,ob05071c,ob05169bat,ob120950b,mb13220b}
However, these measurements have so far been guided by a strategy
focused on the relatively rare events with sufficiently high proper motion
to be resolved by 10m class telescopes, which require at least $\sim 50\,\mas$
separation.  Indeed, \citet{henderson14} specifically advocated such
a strategy.  However, with the prospect of 30m class telescopes, the situation
is now radically altered.  First, the delay time for typical events to
be sufficiently separated is reduced from 10--20 years to 3--5 years.
Second, this implies that almost all planets that have been discovered
or are currently being discovered can be resolved at first 30m AO light,
whether they have typical or fairly slow proper motions, and therefore,
third, are very likely to be resolvable even if there is no proper motion
measurement from the light curve.  Fourth, complete samples of planets
are likely to include many planets without proper-motion 
(so, without $\theta_\e$) measurements \citep{zhu14}, which have been
regarded as less interesting (so, less frequently published).  With AO
observations, these planets would gain host mass and planet mass measurements
of similar quality to those of planets with $\theta_\e$ measurements 
that are derived 
from the light curve.  Thus, the entire field of microlensing planets is likely
to be completely transformed by the advent of 30m AO.  For this reason,
we have given considerable attention to providing information and guidance
to such observations.

\acknowledgments 
This research has made use of the KMTNet system operated by the Korea
Astronomy and Space Science Institute (KASI) and the data were obtained at
three host sites of CTIO in Chile, SAAO in South Africa, and SSO in
Australia.
Work by C.H. was supported by the grant (2017R1A4A101517) of
National Research Foundation of Korea.
Work by H.Y. and W.Z. were partly supported by the National Science Foundation of China (Grant No. 12133005). 
J.C.Y. acknowledges support from U.S.\ N.S.F. Grant No. AST-2108414.
This research uses CFHT data obtained through the Telescope Access Program (TAP), which has been funded by the TAP member institutes.

\input tabnames

\input tab1391

\input tab1253

\input tab1372

\input tab0748

\input tabcmd

\input tab_physall

\clearpage

\begin{figure}
\plotone{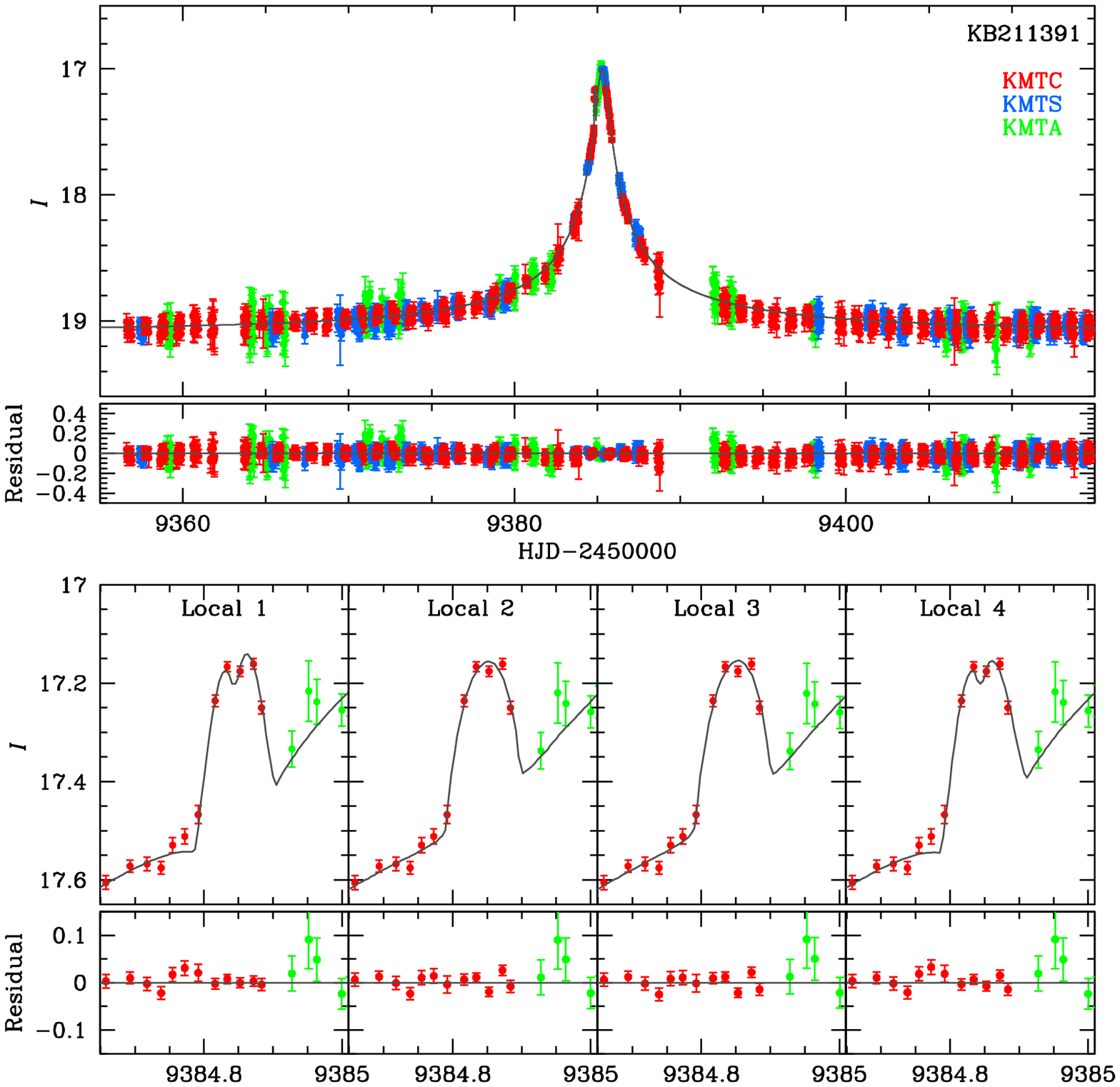}
\caption{Top panel: Light curve and model for KMT-2021-BLG-1391, color-coded
by observatory.  Bottom Panels: Zooms of 0.35 days centered on
the ``bump'' anomaly traced by 6 KMTC points.  The small-$\rho$ 
``double-peaked''  Local 1 is preferred over either of the large-$\rho$
models (Locals 2 and 3)
by $\Delta\chi^2=5$, mainly because it better accounts
for the ``flat top'' or possible ``dimple'' in the bump.  
See Figure~\ref{fig:1391caus}.
}
\label{fig:1391lc}
\end{figure}

\begin{figure}
\plotone{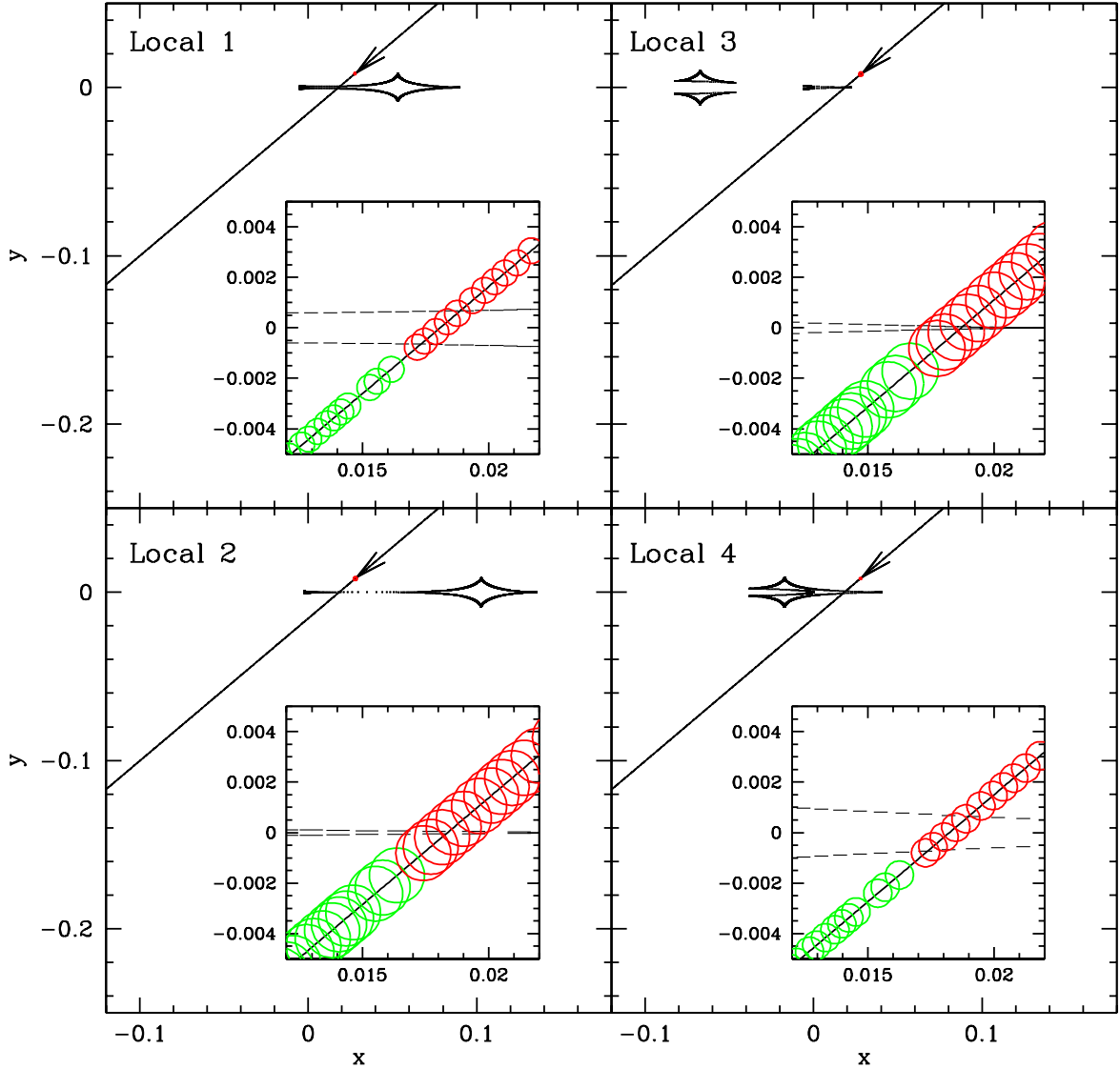}
\caption{Caustic structures for KMT-2021-BLG-1391 for the two ``inner/outer''
pairs of degenerate solutions, i.e., Locals 1 \& 4 and Locals 2 \& 3,
respectively.  The source size is to scale and the epochs
of the observations are shown, with red points from KMTC and green
points from KMTA.  In the first pair of solutions, the source
diameter is slightly smaller than the caustic separation, giving rise
to a ``dimple'' over the peak of the anomaly.  See Figure~\ref{fig:1391lc}.
In the second, the source is much larger than the caustic separation.
The two pairs of solutions predict very different
proper motions and so can ultimately be distinguished by late-time AO imaging.
}
\label{fig:1391caus}
\end{figure}

\begin{figure}
\plotone{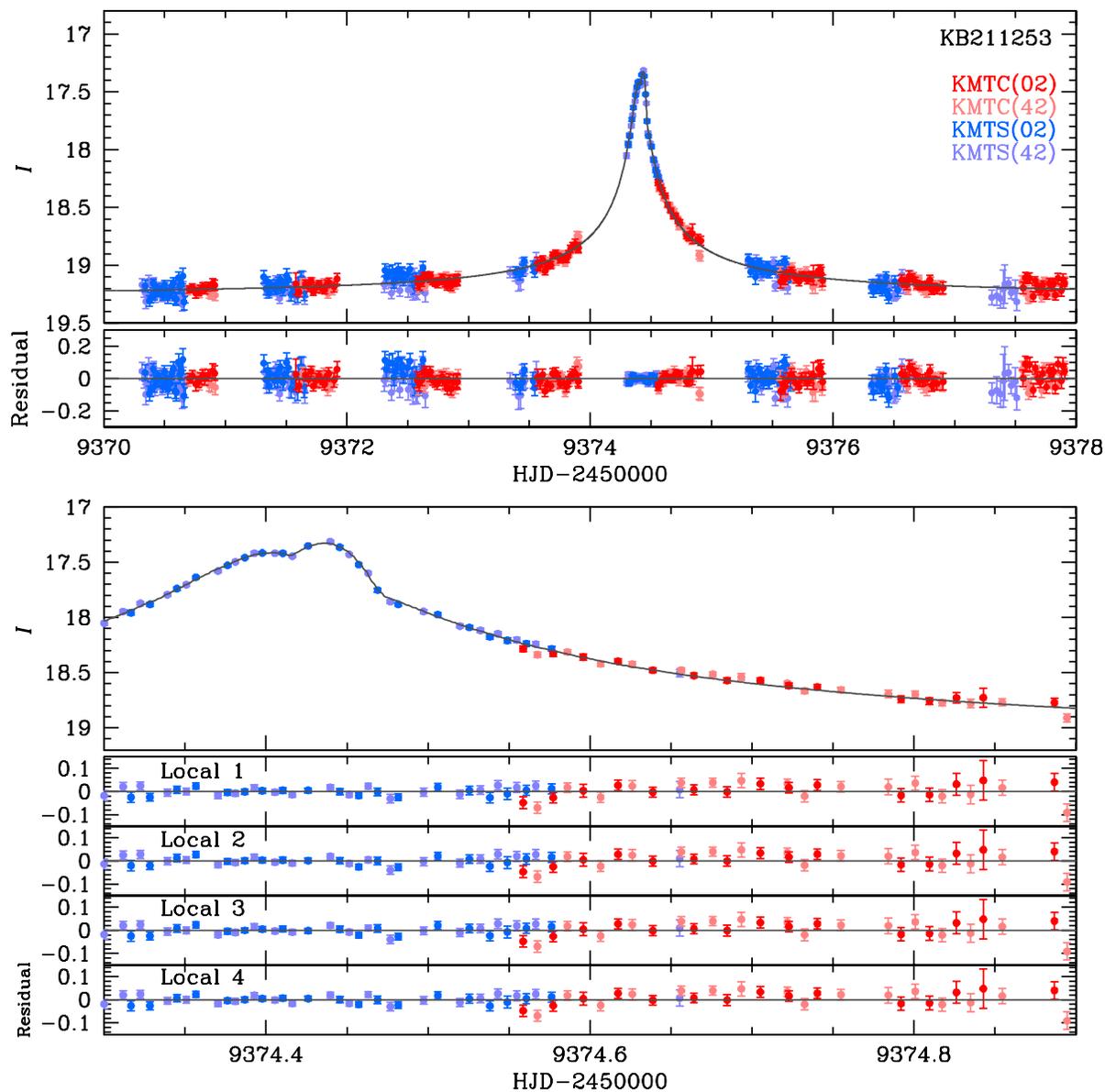}
\caption{Light curve and models for KMT-2021-BLG-1253.  Despite good
coverage of the ``bump'' anomaly, there are four models that fit the
data almost equally well (residuals in lower panels).  However, all
four solutions have similar planet-host mass ratios $q\sim 2.4\times 10^{-4}$
and normalized planet-host separations $s\sim 1$. Nevertheless, they have 
very different normalized source sizes.  See Table~\ref{tab:1253parms} and 
Figure~\ref{fig:1253cau}.}
\label{fig:1253lc}
\end{figure}

\begin{figure}
\plotone{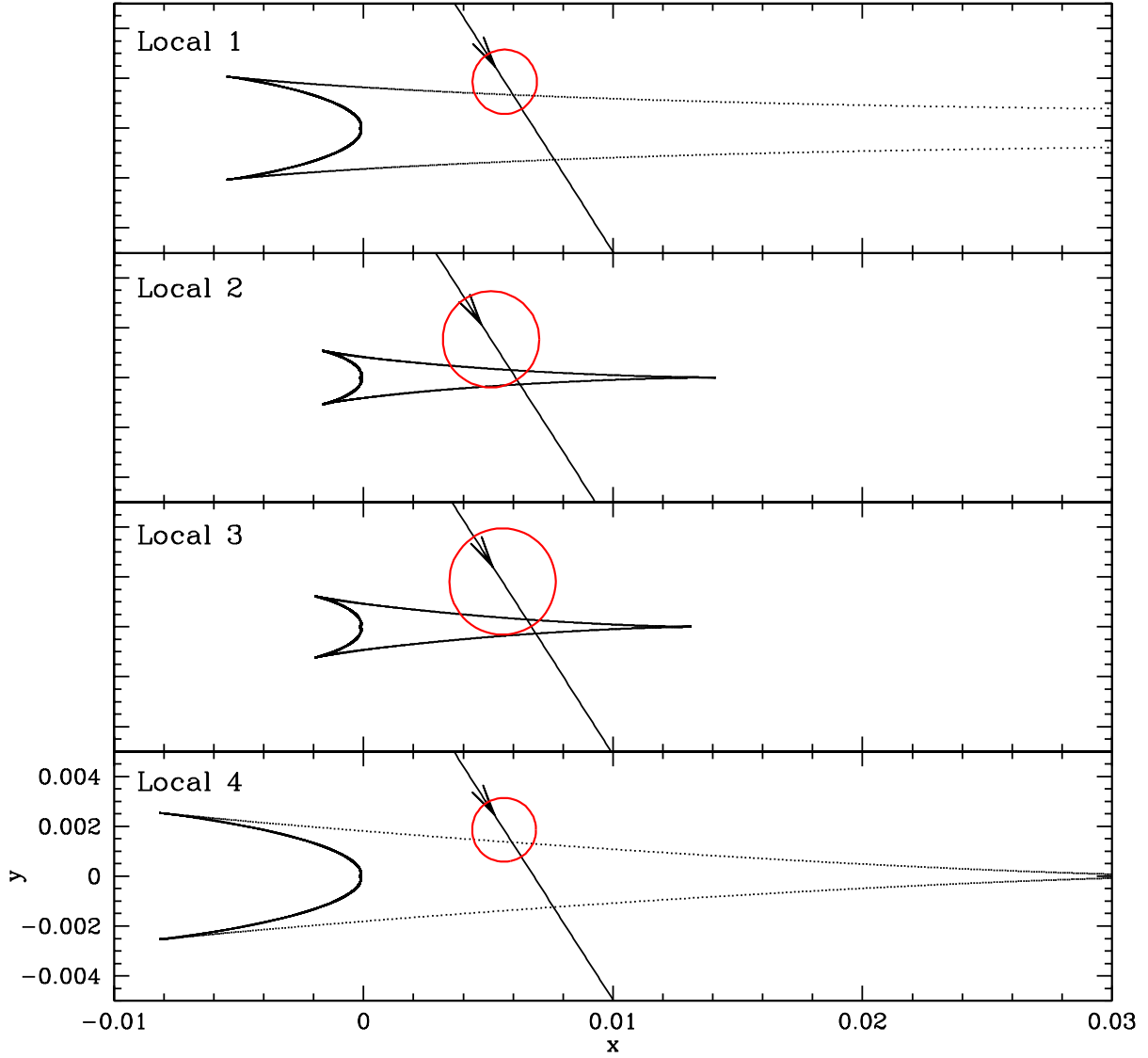}
\caption{Caustics for KMT-2021-BLG-1253.  There are two pairs of solutions:
in Locals 1 and 4, the flat-topped bump is explained by a source whose
diameter is about equal to the separation between two caustic walls,
whereas in Locals 1 and 3, it is explained by a source that is much
larger than this separation (and the caustic is substantially smaller).
Each pair constitutes a classic wide/close degeneracy.  The different
normalized source sizes ($\rho\sim 1.3\times 10^{-3}$ vs.\ 
$\rho\sim 2\times 10^{-3}$) lead to different Bayesian mass and distance 
estimates, but these will ultimately be resolved by AO followup observations.
See Section~\ref{sec:phys-kb211253}.}
\label{fig:1253cau}
\end{figure}

\begin{figure}
\plotone{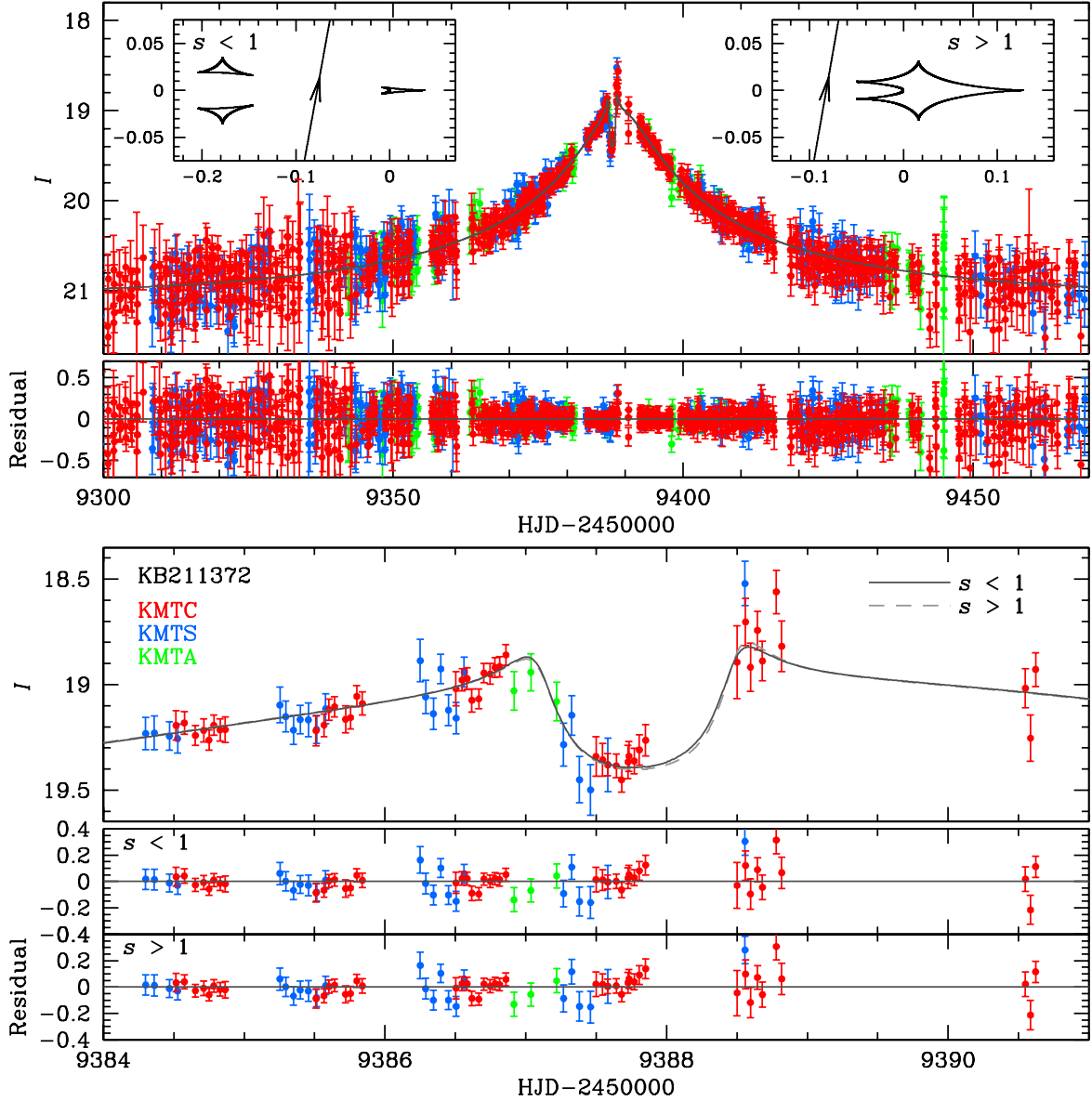}
\caption{Light curve and model for KMT-2021-BLG-1372.
The light curve shows a rounded dip, which is the classic signature
of a minor-image perturbation for which the source misses
the two planetary caustics.  The anomaly is near peak, implying that
the trajectory is roughly perpendicular to the planet-host axis
and that there should be a classic ``inner/outer'' degeneracy \citep{gaudi97}.
These expectations are confirmed by the caustic structures shown in the insets.
}
\label{fig:1372lc}
\end{figure}

\begin{figure}
\plotone{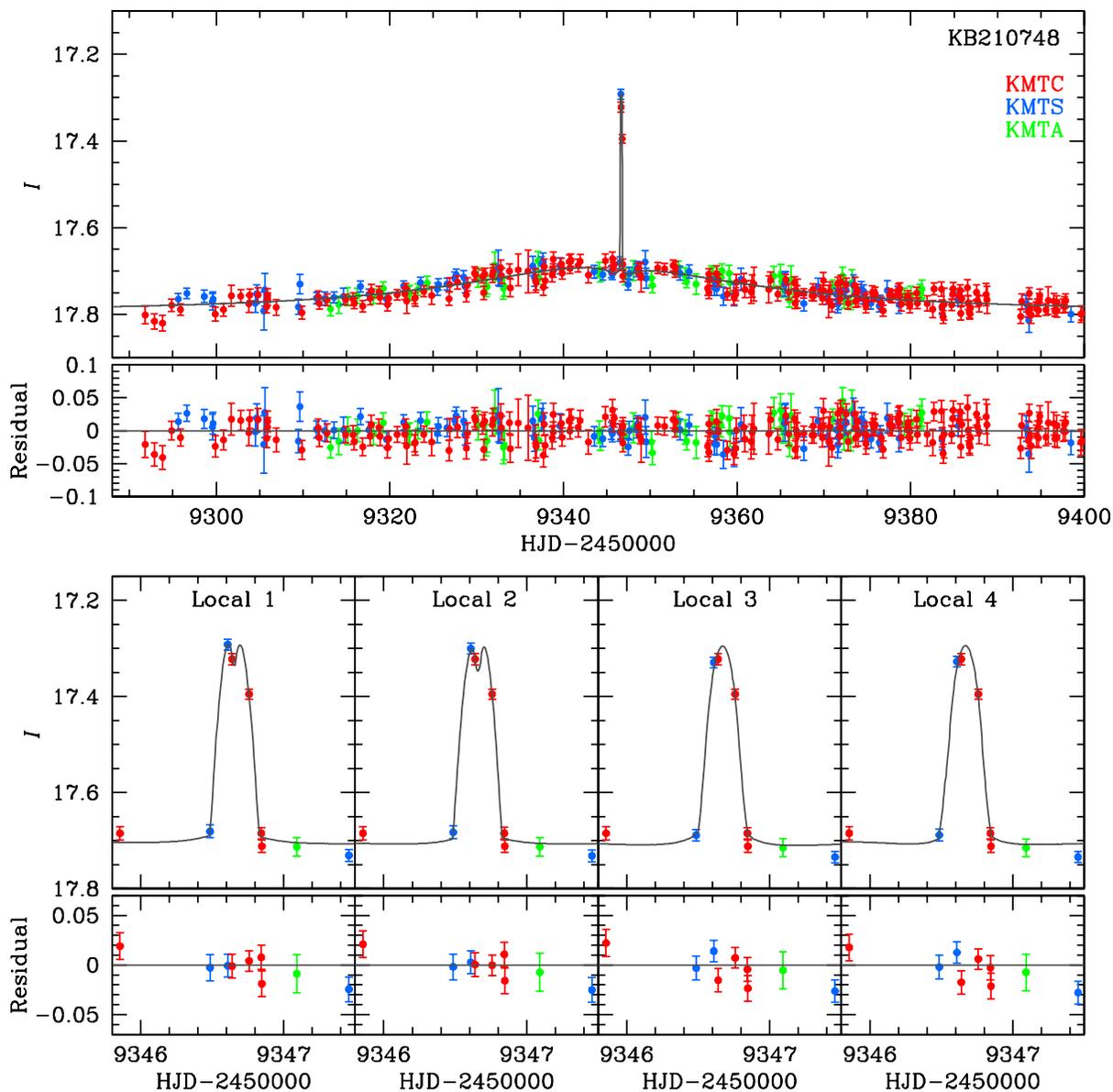}
\caption{Top panel: Light curve and model for KMT-2021-BLG-0748 color-coded
by observatory.  Bottom Panel: Four zooms of 1.7 days centered on the
``bump'' anomaly, which is traced by 3 points.  Locals 1 and 2 are
preferred by $\Delta\chi^2\sim 10$ over Locals 3 and 4 mainly because 
the overall light-curve normalization favors the first point on the
bump (from KMTS) being brighter than the second (from KMTC).  Hence,
the $\chi^2$ difference is not fully reflected in the residuals of the
bump region.  See Figure~\ref{fig:0748cau} for the origin of this difference.
}
\label{fig:0748lc}
\end{figure}

\begin{figure}
\plotone{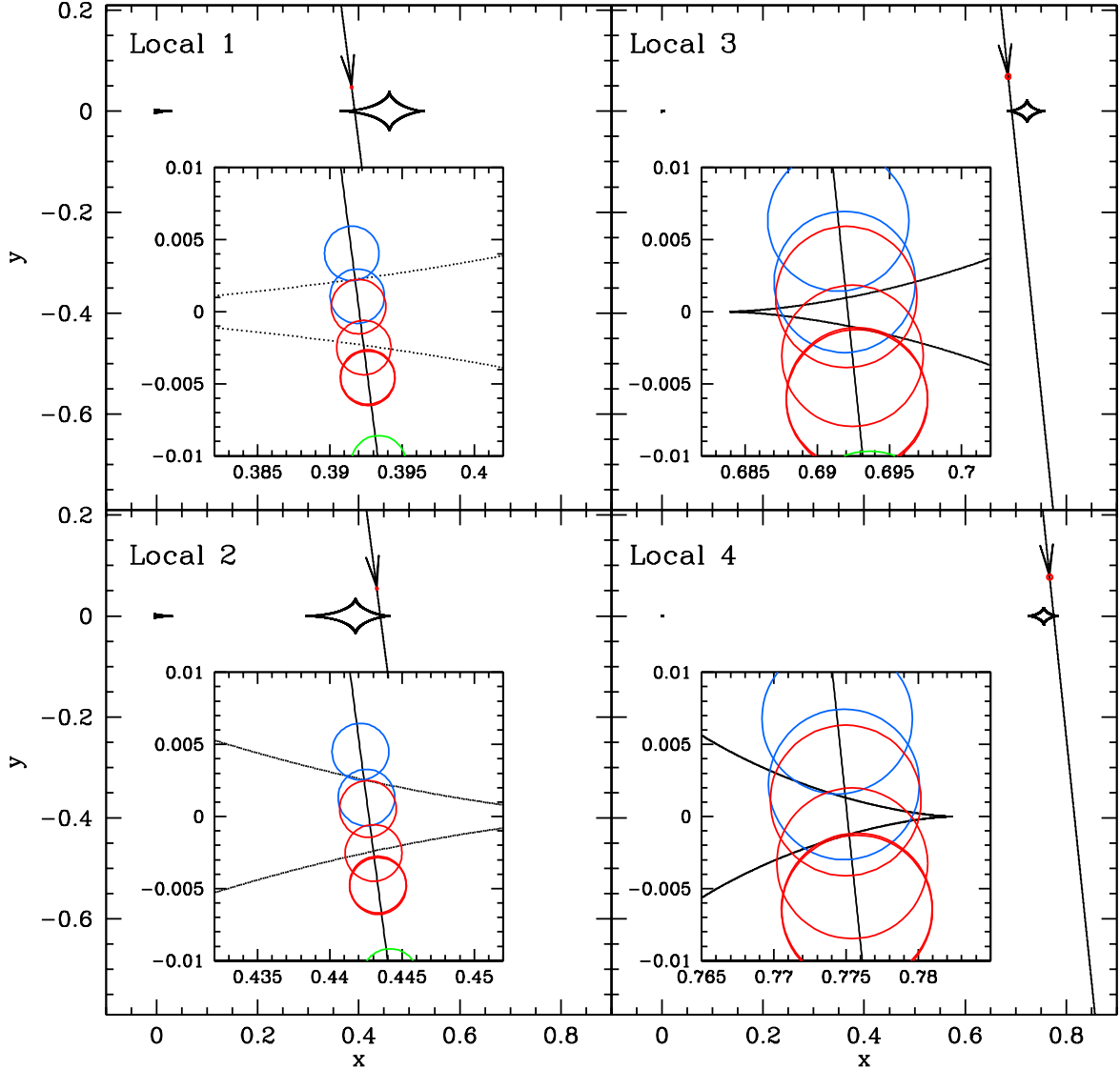}
\caption{Caustic geometries for the four Locals of KMT-2021-BLG-0748.
Locals 1 and 2 are a classic example of the inner/outer degeneracy
\citep{gaudi97}, i.e. the source passes just inside the inner cusp of the
planetary caustic in Local 1 and just inside the outer cusp in Local 2.
The source is smaller than the caustic separation, which induces
``dimples'' in the first two zoom panels of Figure~\ref{fig:0748lc}.
Morphologically, Locals 3 and 4 are ``satellite'' solutions of Locals 1 and 2,
in which the source passes closer to the cusp.  However, the source is about
3 times larger, which induces smooth bumps (i.e., no dimples)
in the last two zoom panels of Figure~\ref{fig:0748lc}.  And this change
induces large changes in $s$ and $q$.
}
\label{fig:0748cau}
\end{figure}

\begin{figure}

\plotone{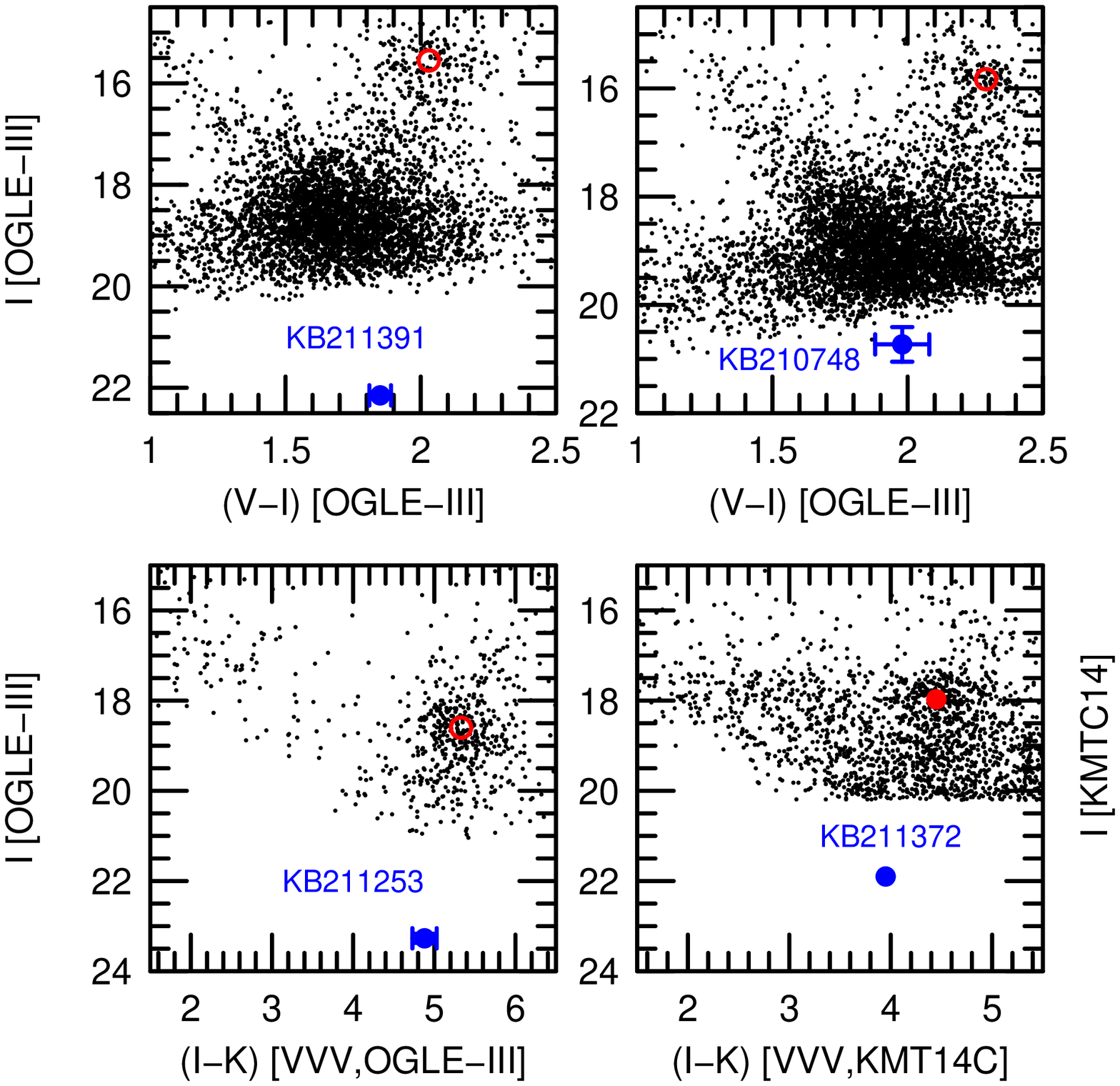}
\caption{CMDs for each of the four planets reported here.
The source positions (blue) and clump-giant centroids (red) are shown
for all events.
}
\label{fig:allcmd}
\end{figure}

\end{document}

%% file: author.tex
\author{\textsc{
Yoon-Hyun Ryu$^{1}$, 
Youn Kil Jung$^{1}$, 
Hongjing Yang$^{2}$, 
Andrew Gould$^{3,4}$, 
Michael D. Albrow$^{5}$, 
Sun-Ju Chung$^{1,6}$, 
Cheongho Han$^{7}$, 
Kyu-Ha Hwang$^{1}$, 
In-Gu Shin$^{1}$, 
Yossi Shvartzvald$^{8}$, 
Jennifer C. Yee$^{9}$, 
Weicheng Zang$^{2}$,
Sang-Mok Cha$^{1,10}$, 
Dong-Jin Kim$^{1}$,
Seung-Lee Kim$^{1,6}$, 
Chung-Uk Lee$^{1,6}$, 
Dong-Joo Lee$^{1}$,
Yongseok Lee$^{1,10}$, 
Byeong-Gon Park$^{1,6}$, 
Richard W. Pogge$^{4}$
} }

\affil{$^{1}$Korea Astronomy and Space Science Institute, Daejon
34055, Republic of Korea}

\affil{$^{2}$ Department of Astronomy and Tsinghua Centre for Astrophysics, 
Tsinghua University, Beijing 100084, China}

\affil{$^{3}$Max-Planck-Institute for Astronomy, K\"{o}nigstuhl 17,
69117 Heidelberg, Germany}

\affil{$^{4}$Department of Astronomy, Ohio State University, 140 W.
18th Ave., Columbus, OH 43210, USA}

\affil{$^{5}$University of Canterbury, Department of Physics and
Astronomy, Private Bag 4800, Christchurch 8020, New Zealand}

\affil{$^{6}$Korea University of Science and Technology, Korea, 
(UST), 217 Gajeong-ro, Yuseong-gu, Daejeon, 34113, Republic of Korea}

\affil{$^{7}$Department of Physics, Chungbuk National University,
Cheongju 28644, Republic of Korea}

\affil{$^{8}$Department of Particle Physics and Astrophysics, 
Weizmann Institute of Science, Rehovot 76100, Israel}

\affil{$^{9}$ Center for Astrophysics $|$ Harvard \& Smithsonian, 60 Garden
St., Cambridge, MA 02138, USA}

\affil{$^{10}$School of Space Research, Kyung Hee University,
Yongin, Kyeonggi 17104, Republic of Korea}

\affil{$^{11}$National Astronomical Observatories, Chinese Academy of Sciences, Beijing 100101, China}




%% file: tabnames.tex
 \begin{deluxetable}{lrrrrrr}
 \tablecolumns{7} \tablewidth{0pc}
 \tablecaption{\textsc{Event Names, Cadences, Alerts, and Locations}}
 \tablehead{\colhead{Name} & 
\colhead{$\Gamma\,({\rm hr}^{-1})$} &
\colhead{Alert Date} &
\colhead{RA$_{\rm J2000}$} &
\colhead{Dec$_{\rm J2000}$} &
\colhead{$l$} &
\colhead{$b$} }
 \startdata
KMT-2021-BLG-1391 & 3.0 & 21 Jun 2021 & 18:02:44.06 & $-28$:03:37.80 & $+2.66$ & $-2.80$\\
\hline
KMT-2021-BLG-1253 & 4.0 & 10 Jun 2021 & 17:50:28.13 & $-29$:16:45.41 & $+0.26$ & $-1.08$\\
\hline
KMT-2021-BLG-1372 & 1.0 & 17 Jun 2021 & 17:37:57.25 & $-28$:08:54.20 & $-0.21$ & $+1.85$\\
\hline
KMT-2021-BLG-0748 & 0.4 & 10 May 2021 & 18:09:41.12 & $-25$:44:47.69 & $+5.44$ & $-3.03$\\
 \enddata
 \label{tab:names}
 \end{deluxetable}


%% file: tab1391.tex
\begin{deluxetable}{lcccc}
\tablecolumns{5} \tablewidth{0pc} \tablecaption{\textsc{Microlens Parameters for KMT-2021-BLG-1391}} 
\tablehead{\colhead{Parameters}
& \colhead{Local 1} & \colhead{Local 2} & \colhead{Local 3} & \colhead{Local 4} } \startdata
  $\chi^2/\rm{dof}$             &4255.157/4249        &4260.261/4249       &4260.229/4249               &4258.723/4249 \\
  $t_0-2459380$                 &5.291 $\pm$ 0.002    &5.291 $\pm$ 0.002   &5.291 $\pm$ 0.002           &5.290 $\pm$ 0.002 \\
  $u_0$                         &0.012 $\pm$ 0.001    &0.012 $\pm$ 0.001   &0.012 $\pm$ 0.001           &0.012 $\pm$ 0.001 \\
  $t_{\rm E}$ $(\rm{days})$     &31.685 $\pm$ 1.457   &31.612 $\pm$ 1.513  &31.791 $\pm$ 1.574          &31.818 $\pm$ 1.448 \\
  $s$                           &1.027 $\pm$ 0.001    &1.056 $\pm$ 0.009   &0.966 $\pm$ 0.008           &0.991 $\pm$ 0.001 \\
  $q$ $(10^{-5})$               &3.619 $\pm$ 0.279    &4.736 $\pm$ 0.629   &$4.635_{-0.549}^{+0.676}$   &3.700 $\pm$ 0.302  \\
  $\log q$ (mean)                &-4.441 $\pm$ 0.034   &-4.323 $\pm$ 0.056  &-4.332 $\pm$ 0.056          &-4.432 $\pm$ 0.036 \\
  $\alpha$ $(\rm{rad})$         &2.437 $\pm$ 0.004    &2.438 $\pm$ 0.004   &2.438 $\pm$ 0.004           &2.438 $\pm$ 0.004 \\
  $\rho$ $(10^{-3})$            &0.535 $\pm$ 0.041    &1.081 $\pm$ 0.068   &1.077 $\pm$ 0.069           &0.573 $\pm$ 0.049 \\
  $I_S$ [KMTC,pySIS]            &21.963 $\pm$ 0.052   &21.961 $\pm$ 0.054  &21.967 $\pm$ 0.056          &21.968 $\pm$ 0.052 \\
  $I_B$ [KMTC,pySIS]            &19.159 $\pm$ 0.003   &19.159 $\pm$ 0.003  &19.159 $\pm$ 0.003          &19.159 $\pm$ 0.003 \\
  $t_*$ $(\rm{hours})$          &0.407 $\pm$ 0.025    &0.821 $\pm$ 0.031   &0.822 $\pm$ 0.031           &0.437 $\pm$ 0.030 \\
\enddata
\label{tab:1391parms}
\end{deluxetable}

%% file: tab1253.tex
\begin{deluxetable}{lcccc}
\tablecolumns{5} \tablewidth{0pc} 
\tablecaption{\textsc{Microlens Parameters for KMT-2021-BLG-1253}} 
\tablehead{\colhead{Parameters}
& \colhead{Local 1} & \colhead{Local 2} & \colhead{Local 3} & \colhead{Local 4} } \startdata
  $\chi^2/\rm{dof}$             &6496.047/6498        &6499.627/6498              &6497.683/6498             &6498.520/6498            \\
  $t_0-2459370$                 &4.410 $\pm$ 0.001    &4.410 $\pm$ 0.001          &4.409 $\pm$ 0.001         &4.410 $\pm$ 0.001        \\
  $u_0$ $(10^{-3})$             &5.641 $\pm$ 0.501    &$5.373_{-0.489}^{+0.600}$  &5.743 $\pm$ 0.510         &5.683 $\pm$ 0.553        \\
  $t_{\rm E}$ $(\rm{days})$     &9.594 $\pm$ 0.841    &10.051 $\pm$ 0.989         &9.479 $\pm$ 0.814         &$9.547_{-0.801}^{+1.025}$        \\
  $s$                           &1.074 $\pm$ 0.004    &$1.150_{-0.019}^{+0.027}$  &$0.873_{-0.020}^{+0.016}$ &0.938 $\pm$ 0.004        \\
  $q$ $(10^{-4})$               &2.311 $\pm$ 0.261    &$2.367_{-0.269}^{+0.334}$  &2.523 $\pm$ 0.302         &2.282 $\pm$ 0.273        \\
  $\log q$ (mean)                      &-3.637 $\pm$ 0.049   &-3.624 $\pm$ 0.055         &-3.598 $\pm$ 0.052        &-3.643 $\pm$ 0.052 \\
  $\alpha$ $(\rm{rad})$         &1.002 $\pm$ 0.010    &1.003 $\pm$ 0.011          &1.002 $\pm$ 0.010         &0.999 $\pm$ 0.011        \\
  $\rho$ $(10^{-3})$            &1.297 $\pm$ 0.145    &2.022 $\pm$ 0.211          &2.141 $\pm$ 0.197         &1.306 $\pm$ 0.166        \\
  $I_S$ [KMTC,pySIS]            &23.185 $\pm$ 0.099   &23.240 $\pm$ 0.111         &23.171 $\pm$ 0.097        &23.179 $\pm$ 0.108 \\
  $I_B$ [KMTC,pySIS]            &19.287 $\pm$ 0.002   &$19.286_{-0.002}^{+0.003}$ &19.288 $\pm$ 0.002        &19.288 $\pm$ 0.003 \\
  $t_*$ (hours)                &$0.297_{-0.017}^{+0.022}$ &0.488 $\pm$ 0.014  &0.487 $\pm$ 0.015 &$0.298_{-0.019}^{+0.027}$ \\
\enddata
\label{tab:1253parms}
\end{deluxetable}

%% file: tab1372.tex
\begin{deluxetable}{lcc}
\tablecolumns{3} \tablewidth{0pc} 
\tablecaption{\textsc{Microlens Parameters for KMT-2021-BLG-1372}} 
\tablehead{\colhead{Parameters}
& \colhead{$s<1$} & \colhead{$s>1$}  } \startdata
  $\chi^2/\rm{dof}$             &1541.940/1541                &1540.605/1541                  \\
  $t_0-2459380$                 &8.758 $\pm$ 0.055            &8.757 $\pm$ 0.057              \\
  $u_0$                         &0.074 $\pm$ 0.011            &0.075 $\pm$ 0.011      \\
  $t_{\rm E}$ $(\rm{days})$     &71.937 $\pm$ 9.161           &70.665 $\pm$ 8.737 \\
  $s$                           &0.917 $\pm$ 0.006            &1.011 $\pm$ 0.008              \\
  $q$ $(10^{-4})$               &4.175 $\pm$ 0.724            &4.419 $\pm$ 0.715      \\
  $\log q$ (mean)               &-3.386 $\pm$ 0.080           &-3.361 $\pm$ 0.075             \\
  $\alpha$ $(\rm{rad})$         &4.896 $\pm$ 0.015            &4.896 $\pm$ 0.015             \\
  $\rho$ $(10^{-4})$            &$<42$    &$<42$     \\
  $I_S$ [KMTC,pyDIA]            &21.919 $\pm$ 0.158           &21.897 $\pm$ 0.156            \\
  $I_B$ [KMTC,pyDIA]            &$21.750_{-0.083}^{+0.105}$   &$21.762_{-0.087}^{+0.107}$             \\
  $t_*$ $(\rm{hours})$          &$0.184_{-0.165}^{+1.322}$    &$0.148_{-0.129}^{+1.180}$      \\
  $t_*$ $(\rm{days})$          &$<0.3$    & $<0.3$      \\
\enddata
\tablecomments{Limits on $\rho$ and $t_*=\rho t_\e$ are at $2.5\,\sigma$.}
\label{tab:1372parms}

\end{deluxetable}

%% file: tab0748.tex
\begin{deluxetable}{lcccc}
\tablecolumns{5} \tablewidth{0pc} 
\tablecaption{\textsc{Microlens Parameters for KMT-2021-BLG-0748}} 
\tablehead{\colhead{Parameters}
& \colhead{Local 1} & \colhead{Local 2} & \colhead{Local 3} & \colhead{Local 4}  } \startdata
  $\chi^2/\rm{dof}$             &654.966/655                 &655.837/655                &665.440/655                &664.390/655        \\
  $t_0-2459340$                 &4.684 $\pm$ 0.408           &4.647 $\pm$ 0.398          &4.685 $\pm$ 0.395          &4.620 $\pm$ 0.384  \\
  $u_0$                         &$0.411_{-0.063}^{+0.120}$   &0.402 $\pm$ 0.060          &0.737 $\pm$ 0.082          &0.760 $\pm$ 0.078  \\
  $t_{\rm E}$ $(\rm{days})$     &39.756 $\pm$ 5.871          &41.116 $\pm$ 4.581         &27.880 $\pm$ 2.193         &27.459 $\pm$ 2.018 \\
  $s$                           &$1.267_{-0.033}^{+0.060}$   &1.190 $\pm$ 0.042          &1.455 $\pm$ 0.051          &1.438 $\pm$ 0.057  \\
  $q$ $(10^{-3})$               &1.261 $\pm$ 0.429           &$1.078_{-0.233}^{+0.304}$  &0.597 $\pm$ 0.179          &0.491 $\pm$ 0.148  \\
  $\log q$ (mean)                &-2.913 $\pm$ 0.155          &-2.968 $\pm$ 0.107         &-3.232 $\pm$ 0.135         &-3.316 $\pm$ 0.134 \\
  $\alpha$ $(\rm{rad})$         &1.451 $\pm$ 0.029           &1.449 $\pm$ 0.025          &1.474 $\pm$ 0.019          &1.473 $\pm$ 0.018  \\
  $\rho$ $(10^{-3})$            &1.523 $\pm$ 0.525           &$1.506_{-0.482}^{+0.356}$  &5.037 $\pm$ 0.535          &5.067 $\pm$ 0.469  \\
  $I_S$ [KMTC,pySIS]            &$20.796_{-0.394}^{+0.244}$  &20.843 $\pm$ 0.226         &19.867 $\pm$ 0.208         &19.814 $\pm$ 0.198 \\
  $I_B$ [KMTC,pySIS]            &$17.862_{-0.014}^{+0.032}$  &17.859 $\pm$ 0.014         &$17.964_{-0.032}^{+0.040}$ &17.973 $\pm$ 0.036 \\
  $t_*$ (hours)                &$1.534_{-0.623}^{+0.305}$    &$1.522_{-0.500}^{+0.278}$      &3.373 $\pm$ 0.197          &3.339 $\pm$ 0.174  \\
\enddata
\label{tab:0748parms}
\end{deluxetable}

%% file: tabcmd.tex
\begin{deluxetable}{lrrrr}
\tablecolumns{6} \tablewidth{0pc}
\tablecaption{\textsc{CMD Parameters for Five 2021 Planets}}
\tablehead{\colhead{Parameter} &
\colhead{KB211391} &
\colhead{KB211253} &
\colhead{KB211372} &
\colhead{KB210748}}
\startdata
$(V-I)_{\rm s}$    & 1.85$\pm$0.04 & N.A.         & N.A.  & N.A.\\ 
$(V-I)_{\rm cl}$   & 2.03$\pm$0.02 & N.A.         & N.A. & 2.29$\pm$0.02\\ 
$(V-I)_{\rm cl,0}$  & 1.06           & 1.06        & 1.06 & 1.06 \\
$(V-I)_{\rm s,0}$   & 0.88$\pm0.05$ & 0.80$\pm$0.10 & 0.75$\pm$0.10& 0.83$\pm$0.10 \\ 
$I_{\rm s}$        & 22.15$\pm$0.05 & 23.27$\pm$0.10 & 21.90$\pm$ 0.16 & 20.73$\pm$0.32\\ 
$I_{\rm cl}$       & 15.55$\pm$0.03 & 18.60$\pm$0.03 & 17.97$\pm$0.05 & 15.83$\pm$0.05\\ 
$I_{\rm cl,0}$     & 14.36          & 14.43         & 14.46          & 14.29\\ 
$I_{\rm s,0}$      & 20.96$\pm$0.06 & 19.10$\pm$0.10 & 18.39$\pm$0.16 & 19.19$\pm$0.32\\  
$\theta_*$ ($\muas$) & 0.244$\pm 0.023$ & 0.517$\pm$0.072 & 0.692$\pm$0.097 & 0.514$\pm$0.089\\ 
\enddata
\tablecomments{Event names are abbreviations for, e.g.,
KMT-2021-BLG-1391.}
\label{tab:cmd}
\end{deluxetable}

%% file: tab_physall.tex
\begin{deluxetable}{lccccccc}
\tablecolumns{8} 
\tablewidth{0pc}\tablecaption{\textsc{Physical properties}} 
\tablehead{\colhead{Event} & \multicolumn{4}{c}{} & \colhead{} &
\multicolumn{2}{c}{Relative Weights}\\
\cline{7-8} \colhead{Models}&\multicolumn{4}{c}{Physical Properties}&  &
\colhead{Gal.Mod.} & \colhead{$\chi^2$}} \startdata
 KB211391  &$M_{\rm host}$ $[M_\sun]$  &$M_{\rm planet}$ $[M_\earth]$  &$D_{\rm L}$ [kpc] &$a_\bot$ [au]\\
 Local 1 &0.39 $\pm$ 0.19  &4.67 $\pm$ 2.25  &$5.55_{-1.65}^{+1.08}$ &$2.41_{-0.65}^{+0.50}$  &&0.919 &1.000\\
 Local 2 &$0.23_{-0.12}^{+0.20}$ &$3.57_{-1.83}^{+3.17}$  &$6.72_{-1.15}^{+0.94}$ &1.63 $\pm$ 0.34  &&0.938 &0.078\\
 Local 3 &$0.23_{-0.12}^{+0.20}$ &$3.52_{-1.80}^{+3.11}$  &$6.72_{-1.15}^{+0.94}$ &1.49 $\pm$ 0.31  &&0.928 &0.079\\
 Local 4 &0.38 $\pm$ 0.19 &4.67 $\pm$ 2.31 &$5.76_{-1.63}^{+1.03}$ &$2.26_{-0.59}^{+0.48}$  &&1.000 &0.168\\
 Total &0.37 $\pm$ 0.19 &4.55 $\pm$ 2.31 &$5.72_{-1.67}^{+1.10}$ &$2.29_{-0.71}^{+0.56}$  && &\\

 \cline{1-8}
 KB211253  &$M_{\rm host}$ $[M_\sun]$  &$M_{\rm planet}$ $[M_\earth]$  &$D_{\rm L}$ [kpc] &$a_\bot$ [au]\\
 Local 1 &$0.290_{-0.15}^{+0.28}$ &$22.23_{-11.79}^{+21.19}$ &$6.33_{-1.35}^{+1.06}$ &1.96 $\pm$ 0.52 &&0.224 &1.000\\
 Local 2 &$0.22_{-0.12}^{+0.24}$ &$17.42_{-9.18}^{+19.08}$  &$6.78_{-1.24}^{+1.00}$ &1.76 $\pm$ 0.42 &&0.937 &0.167\\
 Local 3 &$0.21_{-0.11}^{+0.23}$ &$17.74_{-9.29}^{+19.65}$  &$6.78_{-1.23}^{+0.99}$ &1.30 $\pm$ 0.30 &&1.000 &0.441\\
 Local 4 &$0.29_{-0.15}^{+0.28}$ &$22.05_{-11.71}^{+21.07}$ &$6.32_{-1.36}^{+1.06}$ &1.72 $\pm$ 0.46 &&0.208 &0.290\\
 Adopted &$0.24_{-0.13}^{+0.26}$ &$19.05_{-10.14}^{+20.42}$ &$6.64_{-1.30}^{+1.03}$ &$1.52_{-0.41}^{+0.57}$  && &\\
 \cline{1-8}
 KB211372  &$M_{\rm host}$ $[M_\sun]$  &$M_{\rm planet}$ $[M_\earth]$  &$D_{\rm L}$ [kpc] &$a_\bot$ [au]\\
 $s<1$ &0.42 $\pm$ 0.25  &58.95 $\pm$ 34.51 &$5.97_{-2.41}^{+1.49}$ &2.22 $\pm$ 0.72 &&0.959 &0.513\\
 $s>1$ &0.42 $\pm$ 0.25  &62.41 $\pm$ 36.55 &$5.97_{-2.41}^{+1.49}$ &2.45 $\pm$ 0.79 &&1.000 &1.000\\
 Adopted &0.42 $\pm$ 0.25  &61.27 $\pm$ 35.88 &$5.97_{-2.41}^{+1.49}$ &2.37 $\pm$ 0.77 && &\\
 \cline{1-8}
 KB210748  &$M_{\rm host}$ $[M_\sun]$  &$M_{\rm planet}$ $[M_J]$  &$D_{\rm L}$ [kpc] &$a_\bot$ [au]\\
 Local 1 &$0.36_{-0.20}^{+0.31}$ &$0.47_{-0.26}^{+0.42}$ &$6.09_{-1.31}^{+0.91}$ &2.46 $\pm$ 0.67 &&1.000 &1.000\\
 Local 2 &$0.36_{-0.20}^{+0.32}$ &$0.41_{-0.23}^{+0.36}$ &$6.10_{-1.32}^{+0.91}$ &2.35 $\pm$ 0.64 &&0.772 &0.647\\
 Adopted &$0.36_{-0.20}^{+0.32}$ &$0.45_{-0.25}^{+0.40}$ &$6.09_{-1.31}^{+0.92}$ &2.43 $\pm$ 0.67 && &\\
 \enddata
 \label{tab:physall}
\end{deluxetable}